\documentclass[aps,prb,twocolumn,superscriptaddress]{revtex4-1}

\usepackage{physics}
\usepackage{bbold}
\usepackage{lgrind}        
\usepackage{natbib}    
\usepackage{color}         
\usepackage[pdftex]{graphicx}      
\usepackage{epsf}          
\usepackage{bm}            
\usepackage{amsmath}
\usepackage{thumbpdf}
\usepackage{esint}
\usepackage[belowskip=-10pt,aboveskip=0pt]{caption}
\usepackage{subcaption}
\usepackage[colorlinks=true]{hyperref} 
\usepackage{placeins}
\usepackage{multirow}
\usepackage{multicolumn}
\begin{document}

\title{Origins of Anisotropic Transport in Electrically-Switchable Antiferromagnet $\mathrm{Fe_{1/3}NbS_2}$}


\author{Sophie F. Weber}
\affiliation{Department of Physics, University of California, Berkeley, CA 94720, USA}
\affiliation{Materials Science Division, Lawrence Berkeley National Laboratory, Berkeley, CA 94720, USA}
\author{Jeffrey B. Neaton}
\affiliation{Department of Physics, University of California, Berkeley, CA 94720, USA}
\affiliation{Kavli Energy NanoScience Institute at Berkeley, Berkeley, CA 94720, USA}

\date{\today}
\begin{abstract}
Recent experiments on the antiferromagnetic intercalated transition metal dichalcogenide $\mathrm{Fe_{1/3}NbS_2}$ have demonstrated reversible resistivity switching by application of orthogonal current pulses below its magnetic ordering temperature, making $\mathrm{Fe_{1/3}NbS_2}$ promising for spintronics applications. Here, we perform density functional theory calculations with Hubbard U corrections of the magnetic order, electronic structure, and transport properties of crystalline $\mathrm{Fe_{1/3}NbS_2}$, clarifying the origin of the different resistance states. The two experimentally proposed antiferromagnetic ground states, corresponding to in-plane stripe and zigzag ordering, are computed to be nearly degenerate. In-plane cross sections of the calculated Fermi surfaces are anisotropic for both magnetic orderings, with the degree of anisotropy sensitive to the Hubbard U value. The in-plane resistance, computed within the Kubo linear response formalism using a constant relaxation time approximation, is also anisotropic, supporting a hypothesis that the current-induced resistance changes are due to a repopulating of AFM domains. Our calculations indicate that the transport anisotropy of $\mathrm{Fe_{1/3}NbS_2}$ in the zigzag phase is reduced relative to stripe, consistent with the relative magnitudes of resistivity changes in experiment. Finally, our calculations reveal the likely directionality of the current-domain response, specifically, which domains are energetically stabilized for a given current direction.
\end{abstract}


\maketitle
\section{Introduction}
Due to the bit-like nature of electronic spins, magnetic materials are natural candidates for storage and sensing devices. In particular, the scaling advantages of electrical current over magnetic fields makes spintronic materials whose magnetism can be controlled by current especially desirable\cite{Manchon2019}. The underlying mechanism for current-induced magnetic switching is generally thought to be spin-orbit torque; the applied electric current, in a manner dictated by crystal symmetries, induces a polarization in conduction electrons, thereby creating an effective magnetic field\cite{Manchon2008,Manchon2009,Belkov2008,Fukami2017,Sinova2015,Zelezny2017}. This effective field imparts a torque on the localized magnetic moments, enabling them to switch to different orientations.\\
\indent There has been growing interest in electrically induced switching in antiferromagnetic (AFM) compounds. AFMs have been reported to switch (via a rotation of the N\'{e}el vector) at THz rates by electrical current  compared to a nominal $\sim$ GHz limit for FMs\cite{Olejnik2018}. Moreover, their vanishing bulk magnetization makes them insensitive to stray magnetic fields, enhancing their stability for memory storage relative to ferromagnets (FMs). In spite of their appeal, there are just a few reports of AFM materials which can be electronically manipulated; until very recently the only known examples in single crystal form were the collinear AFMs $\mathrm{CuMnAs}$ and $\mathrm{Mn_2Au}$.\cite{Wadley2016,Bodnar2018} (Additionally, current-driven manipulation of AFMs has also been confirmed in heterostructure devices\cite{Moriyama2018,Chen2018}).\\
 \indent Recently, an electrically switchable AFM was discovered among the magnetically intercalated transition metal dichalcogenides (TMDs), layered compounds in which the magnetic ions are intercalated between the layers. These materials have received attention in the past due to their high tunability; by simply varying the intercalated element, concentration of the intercalant, or base TMD, a wide variety of magnetic and electric ground states are induced\cite{Friend1977,VanLaar1971}. Transport experiments by Nair et al.\cite{Nair2019} demonstrated that one particular case, $\mathrm{Fe_{1/3}NbS_2}$, can be switched between states of high and low resistance by applying orthogonal current pulses. The switching occurs below the N\'{e}el temperature of $49$ K, indicating that the magnetic order is relevant to the changes in resistance.\\
\indent However, the origin of the high and low resistance states has yet to be clarified. It has been hypothesized, based on the results of optical polarimetry measurements, that the resistance change is associated with a current-induced repopulation of three AFM domains\cite{Nair2019,Little2020}, analogously to the current-induced switching observed in $\mathrm{CuMnAs}$\cite{Grzybowski2017}. Little et al.\cite{Little2020} point out that this could occur in theory even if the N\'{e}el vector of $\mathrm{Fe_{1/3}NbS2}$ is fully out of plane. If domain repopulation leads to changes in resistance along a given direction, this will necessarily be reflected in the anisotropy of the electronic structure and transport for a single domain.\\
\indent In what follows, we perform density functional theory (DFT) calculations of the electronic structure and the nature of the magnetic order in $\mathrm{Fe_{1/3}NbS_2}$. We find an AFM ground state, and two nearly degenerate in-plane magnetic orderings corresponding to previously reported ``stripe" and ``zigzag" AFM states. We find that the Fermi surfaces for stripe and zigzag order are both anisotropic in the $k_x$-$k_y$ plane, though the in-plane anisotropy is larger for stripe order. Using our DFT electronic structure and a constant relaxation time approximation within the Kubo linear response formalism, we find that with stripe order the resistivity along the $[120]$ crystallographic axis is roughly twice as large as along the orthogonal $[100]$ direction. On the other hand, the resistivity along $[100]$/ $\hat{x}$ is larger than $[120]$/$\hat{y}$, and the relative anisotropy is reduced for zigzag order. Our computed resistivity tensors for stripe and zigzag order, combined with the experimental switching data, suggest that for both magnetic states a current pulse depopulates the AFM domain whose principle axis is parallel to the current and increases the populations of the other domains.  Our calculations support the domain repopulation hypothesis and provide new insight into the specific current-domain dynamics in $\mathrm{Fe_{1/3}NbS_2}$.
\section{Methods}\label{sec:methods}
For our first-principles density functional theory (DFT) calculations on $\mathrm{Fe_{1/3}NbS_2}$, we employ the Vienna \emph{ab intitio} simulation package (VASP)\cite{Kresse1996} with generalized gradient approximation (GGA) using the Perdew-Burke-Ernzerhof (PBE) functional\cite{Perdew1996} and projector augmented-wave (PAW) method\cite{Blochl1994}. For all DFT calculations we include spin orbit coupling (SOC), and treat it self-consistently. We take $3\mathrm{d}$ and $4\mathrm{s}$; $4\mathrm{p}$, $4\mathrm{d}$, and $5\mathrm{s}$; and $3\mathrm{s}$ and $3\mathrm{p}$ electrons explicitly as valence for $\mathrm{Fe}$, $\mathrm{Nb}$, and $\mathrm{S}$, respectively. We use an energy cutoff of 650 eV for our plane wave basis set. For our $\mathbf{k}$-point grid we use a $\Gamma$-centered mesh of $12\times7\times6$ for the $1\times \sqrt{3}\times1$ orthorhombic supercell consistent with stripe order, and a $6\times7\times6$ mesh for the $2\times\sqrt{3}\times1$ supercell consistent with zigzag AFM order. We use the tetrahedron method\cite{Blochl1994a} for Brillouin zone integrations. These parameters lead to total energy convergence of $<1$ meV/$\mathrm{Fe}$ ion. We use the experimental lattice constants of $a=5.76$ \AA {} and $c=12.20$ \AA, {} and experimental atomic coordinates\cite{VanLaar1971}, having checked that relaxation changes parameters and atomic positions negligibly. For calculations of two-dimensional fermi surfaces and velocities, we use Wannier interpolation as implemented in the post-processing utility postw90 for Wannier90\cite{Marzari1997,Mostofi2014,Yates2007}. We use $208$ and $416$ bands for stripe and zigzag order respectively in our Wannierizations. We select $\mathrm{Fe}$ $\mathrm{d}$, $\mathrm{Nb}$ $\mathrm{d}_{z^2}$, and $\mathrm{S}$ $\mathrm{p}$ orbitals as our localized projections. Cross sections of the Fermi surfaces and Fermi velocities are evaluated on a $k_x\times k_y\times k_z$ grid of $251\times251\times1$. Fermi surface cross sections shown in the the Supplement\cite{suppmat} without band velocities were generated using WannierTools\cite{Wu2018}. The evaluation of the Kubo formula for conductivity is performed using the Wannier-linear-response code\cite{Zelezny2018}. The code calculates linear response properties within the Kubo formalism based on DFT-parameterized tight-binding Hamiltonians, taking the overlap of Wannier functions as input. We use a converged k-grid of $400\times400\times400$ for evaluation of the conductivities.\\
\indent To approximately account for the localized nature of the $\mathrm{Fe}$ $\mathrm{d}$ electrons we add a Hubbard U correction\cite{Anisimov1997}, and we select the rotationally invariant implementation by Dudarev et al.\cite{Dudarev1998}.  We note here that our quantitative results for energetics, Fermi surface cross sections, and transport tensors are highly sensitive to the specific value of Hubbard $\mathrm{U}$ chosen. The Hubbard $\mathrm{U}$, an ad-hoc parameter, acts here explicitly on the $\mathrm{Fe}$ $\mathrm{d}$ states, which have a very large weight near the Fermi energy in $\mathrm{Fe_{1/3}NbS_2}$; therefore, small changes in $\mathrm{U}$ have a disproportionate effect on bands in an energy window relevant for transport properties (see Supplement for orbital-projected band structures\cite{suppmat}). Given the limitations of PBE+U, to gain confidence in consistent qualitative features in transport anisotropy we perform and describe PBE+U calculations using two different $\mathrm{U}$ values in the main text. We first use PBE+U with $\mathrm{U}=0.3$, $\mathrm{eV}$ following previous work, which results in a magnetoanisotropy energy (MAE) consistent with experiment\cite{Haley2020a}. However, as we noted in Reference \citenum{Haley2020a}, $\mathrm{U}=0.3$ $\mathrm{eV}$ overestimates Heisenberg exchange constants as compared to experiment by several orders of magnitude. This motivates our consideration of a larger value $\mathrm{U}=0.9$ $\mathrm{eV}$ for comparison, which results in smaller (though still overestimated) Heisenberg exchange constants due to increased localization, and also gives the correct sign for the MAE (easy axis along $\mathrm{c}$) while the magnitude of the MAE is overestimated. We note here that, as shown in the Supplement, even if we use a much larger $\mathrm{U}=4$ $\mathrm{eV}$ which gives an incorrect sign for the MAE, the qualitative trends for transport with both zigzag and stripe magnetism are identical to those presented here using $\mathrm{U}=0.3$ $\mathrm{eV}$ and $\mathrm{U}=0.9$ $\mathrm{eV}$, giving us further confidence in the robustness of our results. We refer the reader to the Supplement for further details and discussion\cite{suppmat}.\\
\begin{figure}
\includegraphics[width= \columnwidth]{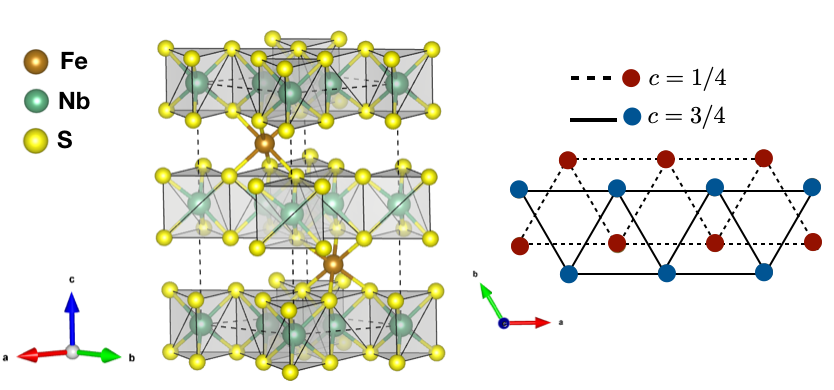}
\caption{\label{fig:struct}Left: hexagonal crystal structure of $\mathrm{Fe_{1/3}NbS_2}$, with space group $P6_322$.The primitive cell contains two $\mathrm{Fe}$ atoms sandwiched between the layers of $\mathrm{NbS_2}$ at $c=1/4$ and $c=3/4$. Right: c-oriented view of the two $\mathrm{Fe}$ layers with ions in layer $c=1/4$ and $c=3/4$ colored red and blue respectively. }
\end{figure}
\section{Crystal Structure} 
$\mathrm{Fe_{1/3}NbS_2}$ is a layered compound with $\mathrm{Fe}$ intercalated between 2H-type TMD $\mathrm{NbS_2}$ layers\cite{Friend1977}. The primitive non-magnetic unit cell is depicted in Figure \ref{fig:struct}. The $\mathrm{Nb}$ atoms are surrounded by the $\mathrm{S}$ atoms in a trigonal prismatic coordination. $\mathrm{Fe_{1/3}NbS_2}$ takes up the space group $P6_322$ [182]. The $\mathrm{Fe}$ atoms are sandwiched between the $\mathrm{NbS_2}$ layers at relative coordinates $(1/3,2/3,1/4)$ and $(2/3,1/3,3/4)$ (Wyckoff position 2d). There are two different $\mathrm{Fe}$ layers stacked along $\mathrm{c}$, with each layer forming a triangular lattice in the a-b plane (note that the a-b plane is what we refer to as ``in-plane" in what follows).\\
\section{Magnetic Order}
\begin{figure}
\begin{subfigure}[t]{0.50\columnwidth}\caption{}\label{fig:stripe}\includegraphics[width=\columnwidth]{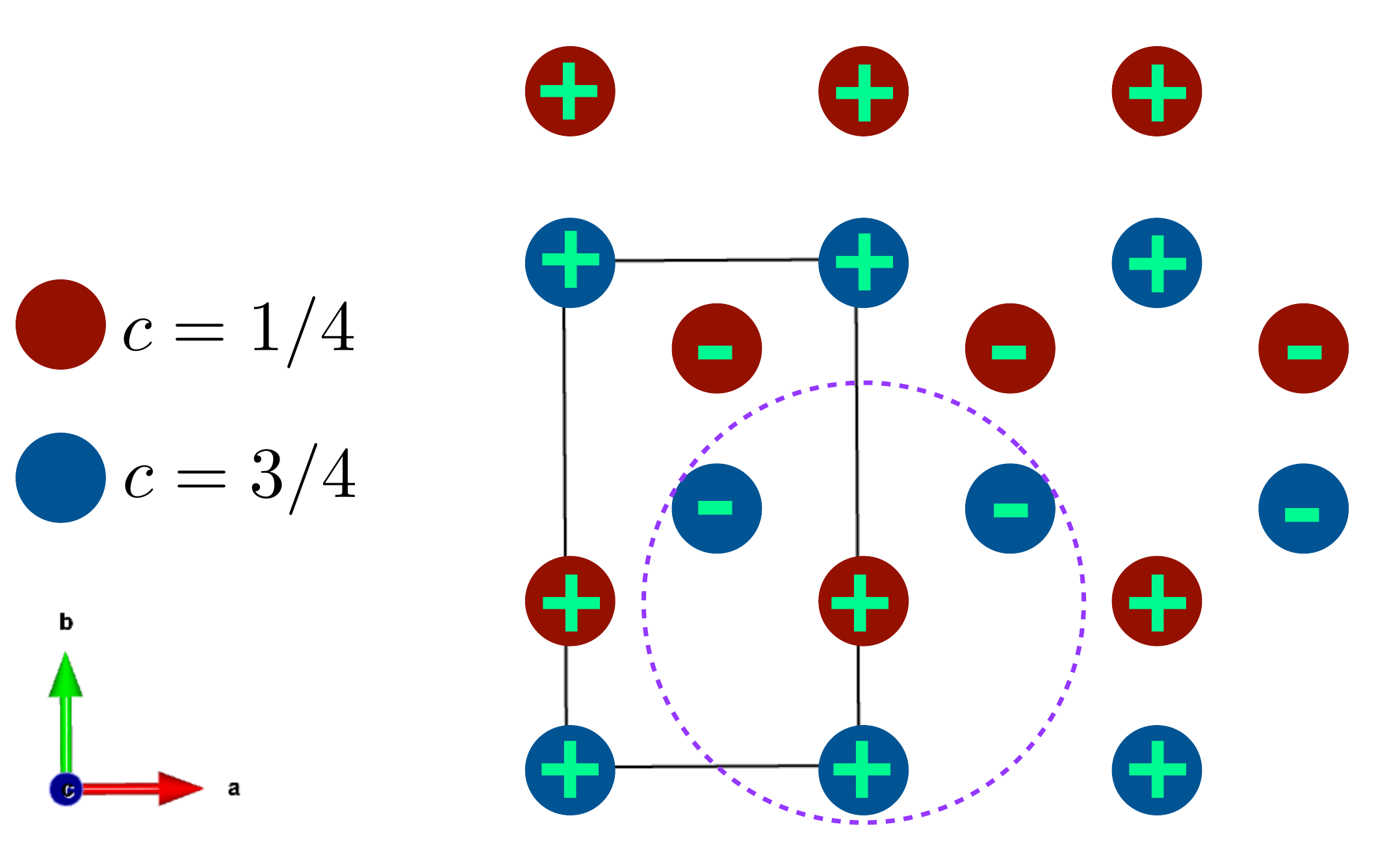}
\end{subfigure}
\begin{subfigure}[t]{0.47\columnwidth}\caption{}\label{fig:zigzag}\includegraphics[width=\columnwidth]{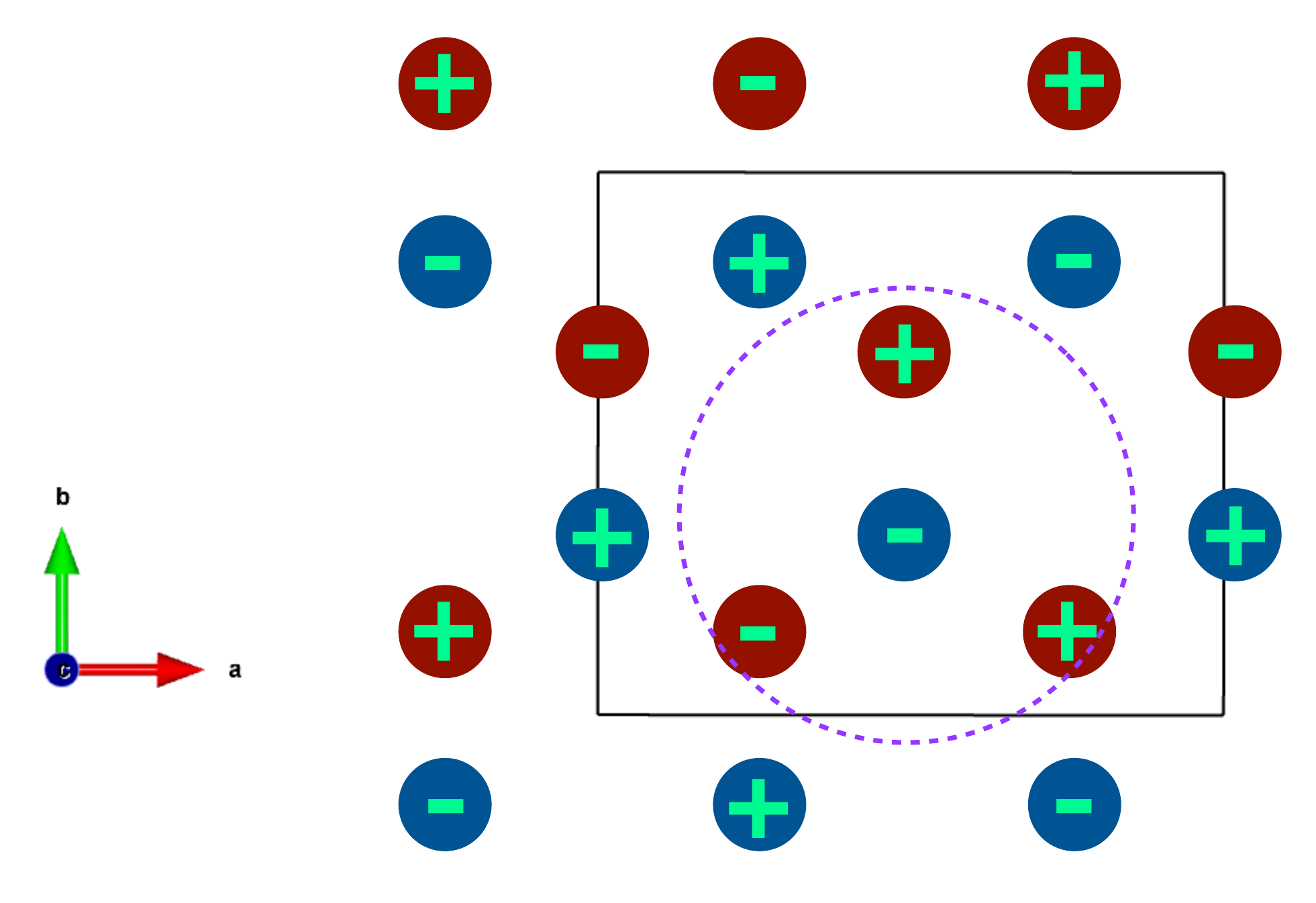}
\end{subfigure}
\caption{Experimentally proposed magnetic orderings, (a) $P_C2_12_12_1$ (a-stripe); (b) $P_c2_12_12$ (a-zigzag), with only $\mathrm{Fe}$ spins shown. In our DFT calculations the N\'{e}el vector is purely out of plane; $+$ and $-$ symbols refer to up and down spins respectively. Magnetic supercells are outlined in black. The orthohexagonal supercell for stripe order in terms of the primitive hexagonal lattice vectors $\mathrm{a}$ and $\mathrm{c}$ is $\mathrm{a}\times\sqrt{3}\mathrm{a}\times\mathrm{c}$ and the supercell for zigzag order is $2\mathrm{a}\times\sqrt{3}\mathrm{a}\times\mathrm{c}$. Dashed purple circles show the three interplanar nearest neighbors for a given ion, which determine whether the planes are ``FM" coupled or ``AFM" coupled; the coupling is AFM in both cases.}\label{fig: mag_orders} 
\end{figure}

The magnetic ground state of $\mathrm{Fe_{1/3}NbS_2}$ is known to be AFM below about $50\mathrm{K}$\cite{Friend1977}, but the nature of the AFM order is highly sensitive to small changes in $\mathrm{Fe}$ concentration. Seminal work more than 40 years ago\cite{VanLaar1971} indicated that for $\mathrm{Fe_xNbS_2}$ with $x=0.323$, an in-plane``zigzag" AFM order of the $\mathrm{Fe}$ spins, with the N\'{e}el vector oriented out of plane along [001] and the spins along one in-plane $\mathrm{Fe}$ bond direction alternating between up and down and between``up up" and ``down down" along the other two bond directions (Figure \ref{fig:zigzag}). However, another neutron scattering study by Suzuki et al.\cite{Suzuki1993} with $x=0.297$ found evidence for a stripe AFM ground state, with rows of spins along one $\mathrm{Fe}$ bond direction alternating between all up and all down (Figure \ref{fig:stripe}).\\ 
 \indent We perform DFT calculations for both experimentally proposed collinear magnetic orderings, with the N\'{e}el vector taken along $\mathrm{c}$, corresponding to magnetic space groups $P_C2_12_12_1$ (stripe)\cite{Suzuki1993} and $P_c2_12_12$ (zigzag)\cite{VanLaar1971}  (see figure \ref{fig: mag_orders}). In what follows, we will refer to them as a-stripe and a-zigzag respectively, with the ``a" indicating that adjacent planes of $\mathrm{Fe}$ ions are AFM coupled. From our PBE+U calculations, these two magnetic orders at the stoichiometric $\mathrm{Fe}$ concentration of $x=\frac{1}{3}$ are nearly degenerate; the energy differences between the magnetic states are $0.9$ and $2.5$ $\mathrm{meV}$ per $\mathrm{Fe}$ atom for $\mathrm{U}=0.3$ and $\mathrm{U}=0.9$ $\mathrm{eV}$, respectively. Additionally, the slightly preferred ground state switches from a-stripe for $\mathrm{U}=0.3$ $\mathrm{eV}$ to a-zigzag for $\mathrm{U}=0.9$ $\mathrm{eV}$. \\
\indent The near-degeneracy of a-stripe and a-zigzag phases can be understood quantitatively from a Heisenberg Hamiltonian also discussed in Reference \citenum{Haley2020a} for PBE+U calculations with $\mathrm{U}=0.3$ $\mathrm{eV}$. We return to it here and discuss the exchange constants in the case of both $\mathrm{U}=0.3$ $\mathrm{eV}$ and $\mathrm{U}=0.9$ $\mathrm{eV}$. Neglecting the antisymmetric spin exchange constants which could lead to slight deviations from fully collinear order, magnetic contributions to the energy of $\mathrm{Fe_{1/3}NbS_2}$ can be described approximately by the following Heisenberg Hamiltonian for the $\mathrm{Fe}$ lattice:
\begin{multline}
H=E_0+\sum_{\langle ij \rangle}J_1S^2+\sum_{\langle \langle ij \rangle \rangle}J_2S^2+\sum_{{\langle ij \rangle}_c}J_{1c}S^2 \\
+\sum_{{\langle \langle ij \rangle \rangle}_c}J_{2c}S^2+\sum_{{\langle \langle \langle ij \rangle \rangle \rangle}_c}J_{3c}S^2-\sum_iD(S_i^z)^2,
 \label{eq:HeisHam}
 \end{multline}
 where $S=2$ is the spin value of $\mathrm{Fe}^{2+}$; one, two and three pairs of brackets distinguish Heisenberg exchange constants between equidistant nearest, next-nearest and third-nearest neighbors respectively; and the $c$ subscript refers to interplanar, rather than in-plane couplings. The last term is the magnetoanisotropy energy (MAE) which, while relevant to our studies in Reference \citenum{Haley2020a}, we neglect here as both a-stripe and a-zigzag phases have their N\'{e}el vectors fully along [001]. $E_0$ encompasses nonmagnetic contributions to the energy. Note that we neglected the third nearest neighbor exchange $J_{3c}$ in Reference \citenum{Haley2020a} as it did not qualitatively alter our conclusions. To obtain the five coupling constants plus $E_0$ we fit our DFT total energies for six inequivalent collinear magnetic configurations (discussed in the Supplement\cite{suppmat}), which include the a-stripe and a-zigzag phases, to Equation \ref{eq:HeisHam} for each $\mathrm{U}$ value studied.\\
 \indent We find for both sets of PBE+U calculations that the in-plane and interplanar nearest neighbor exchange constants $J_1$ and $J_{1c}$ are antiferromagnetic ($J>0$) and significantly larger in magnitude than the other three exchange constants $J_2$, $J_{2c}$ and $J_{3c}$ (which are all ferromagnetic, $J<0$). We note that this is also qualitatively consistent with a previous DFT study of the exchange constants in $\mathrm{Fe_{1/3}NbS_2}$ with no Hubbard $\mathrm{U}$ correction ($\mathrm{U}=0$ $\mathrm{eV}$)\cite{Mankovsky2016}. Focusing on the experimentally relevant a-stripe and a-zigzag phases, the difference in energy between a-stripe and a-zigzag phase using the above equation is given by
 \begin{equation}
 E_{a-stripe}-E_{a-zigzag}=4J_{2c}S^2-4J_2S^2-8J_{3c}S^2,
 \label{eq:stripe_zigdiff}
 \end{equation}
 where again, the interplanar $J_{2c}$,  $J_{3c}$ and in-plane $J_2$ are all FM ($J<0$). We see then that the condition for the a-stripe phase to be favored is $\abs{J_{2c}}>\abs{J_2}+2\abs{J_{3c}}$, whereas the a-zigzag is energetically favored when $\abs{J_{2c}}<\abs{J_2}+2\abs{J_{3c}}$. Thus, the fact that the ground state changes from a-stripe to a-zigzag phase as a function of U can be connected to a shift in calculated relative values of three very small exchange constants (a table with all Heisenberg exchange constants in equation \ref{eq:HeisHam} for both $\mathrm{U}$ values is provided in the Supplement\cite{suppmat}). Specifically, while the magnitudes of most of the $\mathrm{U}=0.9$ $\mathrm{eV}$ exchange constants diminish fairly uniformly relative to those calculated with $\mathrm{U}=0.3$ $\mathrm{eV}$ (as expected due to increased electron localization with larger $\mathrm{U}$), the in-plane next-nearest neighbor exchange constant $J_2$ grows with $\mathrm{U}$. This is likely due the enhanced hybridization between $\mathrm{Nb}$ $\mathrm{d}$ and $\mathrm{Fe}$ $\mathrm{d}$ states in the $k_z=0$ plane for PBE+U with $\mathrm{U}=0.9$ $\mathrm{eV}$ compared to $\mathrm{U}=0.3$ $\mathrm{eV}$ (see orbital projected band structures in Supplement\cite{suppmat}). Because the magnetism in $\mathrm{Fe_{1/3}NbS_2}$ and other magnetically intercalated TMDs is likely RKKY-mediated\cite{Friend1977}, enhanced hybridization between $\mathrm{Fe}$ and $\mathrm{Nb}$ states in the $k_z=0$ plane would be consistent with larger long-range in-plane couplings.\\
 \indent Direct conclusions regarding the magnetic ground state of $\mathrm{Fe_{1/3}NbS_2}$ for intercalations slightly below or above $x=\frac{1}{3}$ cannot, strictly speaking, be made from our PBE+U calculations using this stoichiometric intercalation. Nevertheless, our PBE+U result of competing ground states at $x=\frac{1}{3}$ is consistent with the experimental sensitivity of the magnetic ground state to small deviations from $\frac{1}{3}$. Moreover, the change in our computed exchange constants, and consequently in the magnetic ground state, for small changes in the $\mathrm{U}$ parameter are consistent with the unpublished neutron scattering report\cite{Wu2021} suggesting that a-stripe and a-zigzag phases may coexist at $x=\frac{1}{3}$. If the experimental ground state at $x=\frac{1}{3}$ is in fact a superposition of a-stripe and a-zigzag phases, the changes in magnetic energetics as a function of $\mathrm{U}$ could reflect the fact that this compound is incompletely described by single set of Heisenberg exchange constants.  In any case, the experimental relevance of the a-stripe and a-zigzag phases, in addition to our PBE+U findings that they are energetically competitive, motivate us to study the transport anisotropy of both magnetic orders in what follows.\\
\section{Fermi Surface Cross Sections} 
\begin{figure*}
\begin{subfigure}[t]{0.48\columnwidth}\caption{}\label{fig:Vx_stripe_Up3}\includegraphics{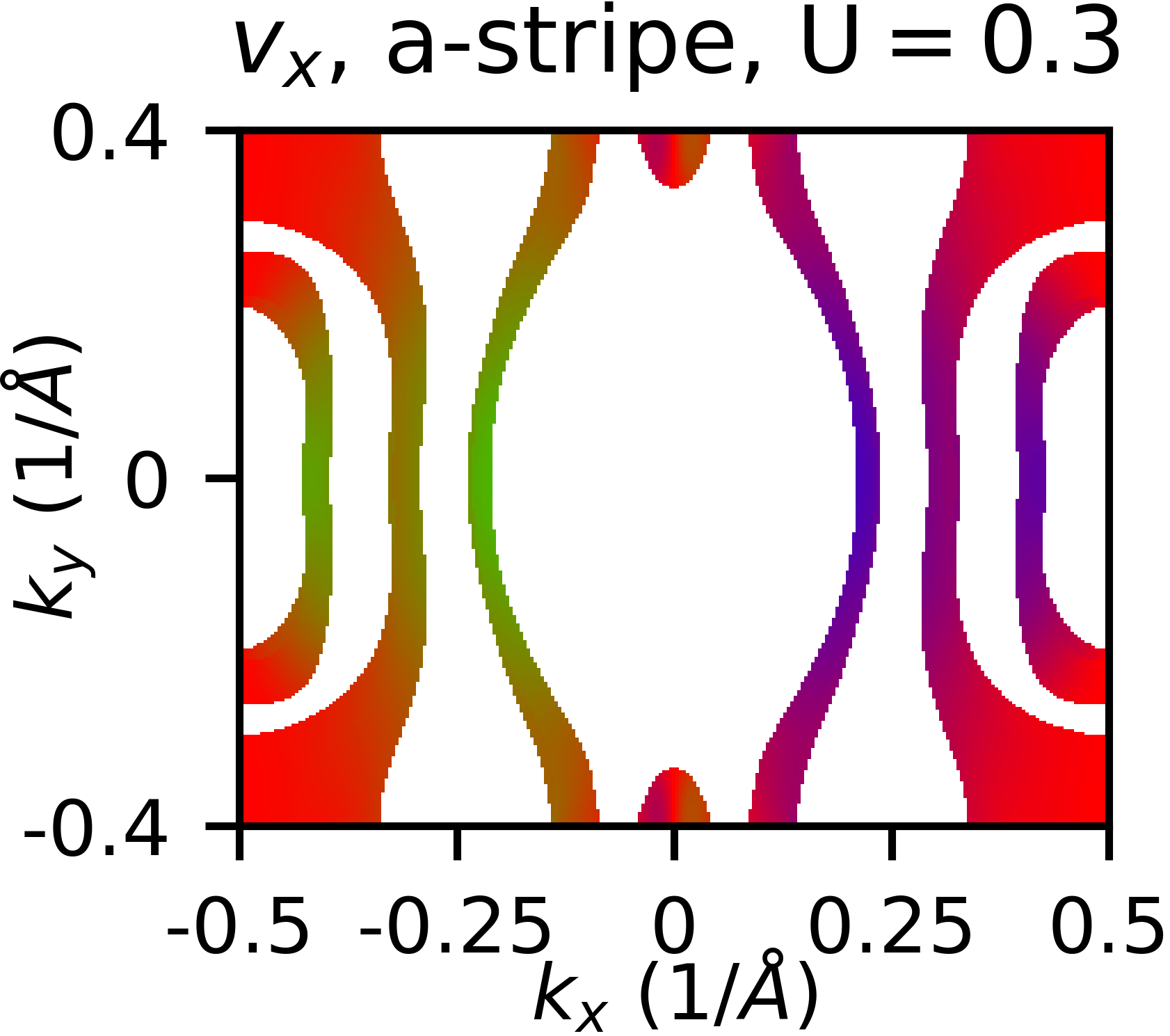}
\end{subfigure}
\begin{subfigure}[t]{0.48\columnwidth}\caption{}\label{fig:Vy_stripe_Up3}\includegraphics{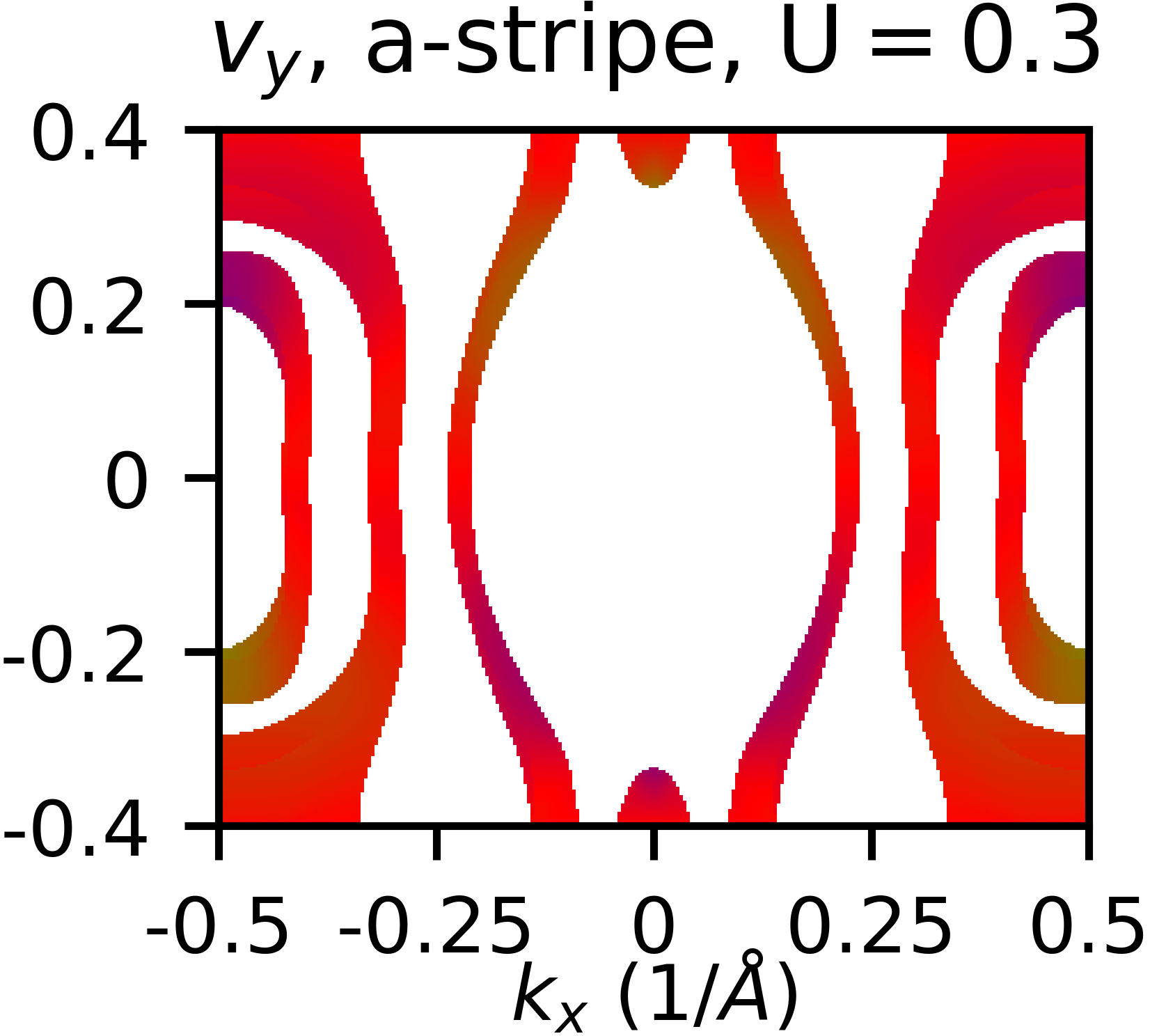}
\end{subfigure}
\begin{subfigure}[t]{0.48\columnwidth}\caption{}\label{fig:Vx_zig_Up3}\includegraphics{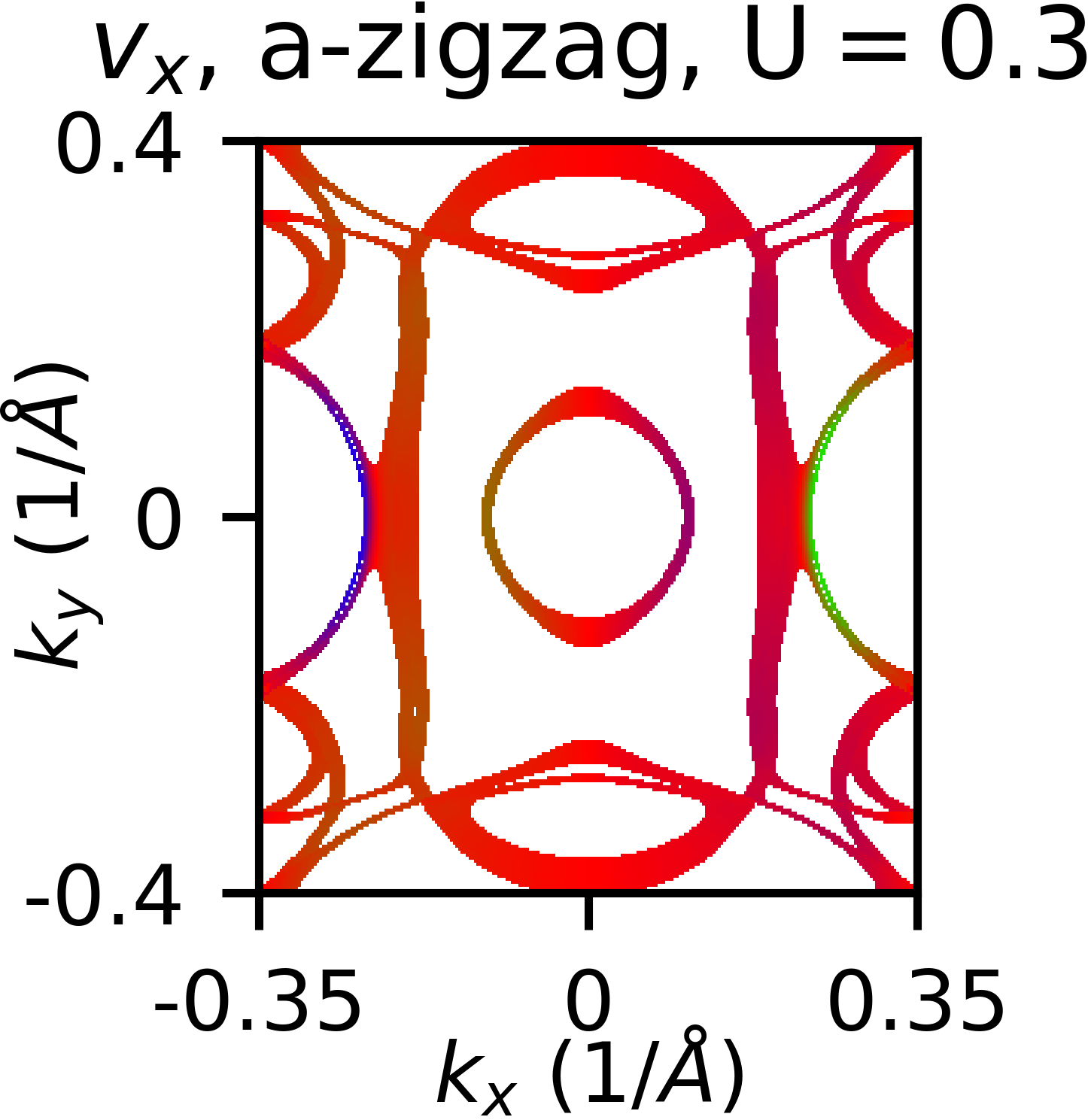}
\end{subfigure}
\begin{subfigure}[t]{0.48\columnwidth}\caption{}\label{fig:Vy_zig_Up3}\includegraphics{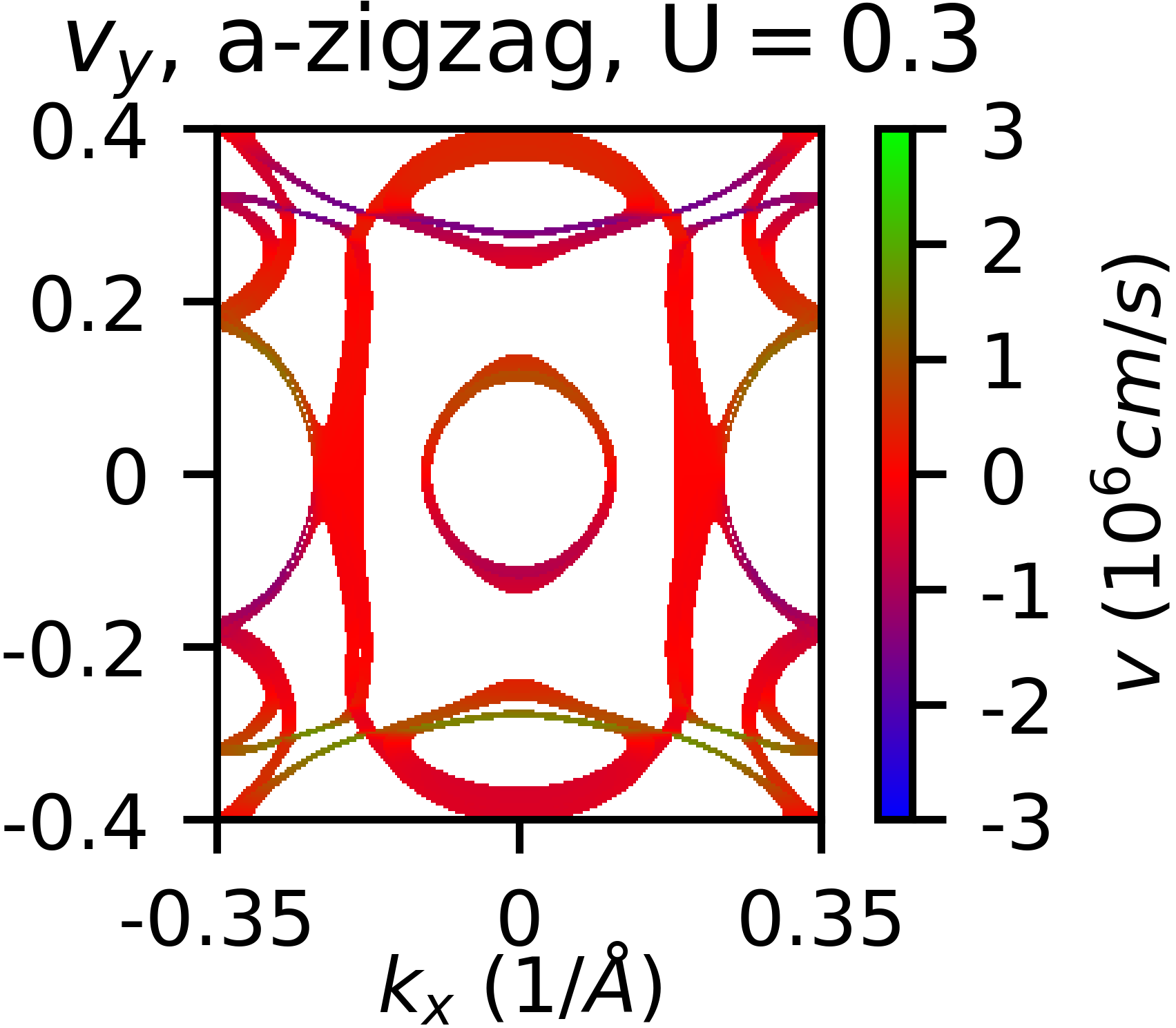}
\end{subfigure}
\begin{subfigure}[t]{0.48\columnwidth}\caption{}\label{fig:Vx_stripe_Up9}\includegraphics{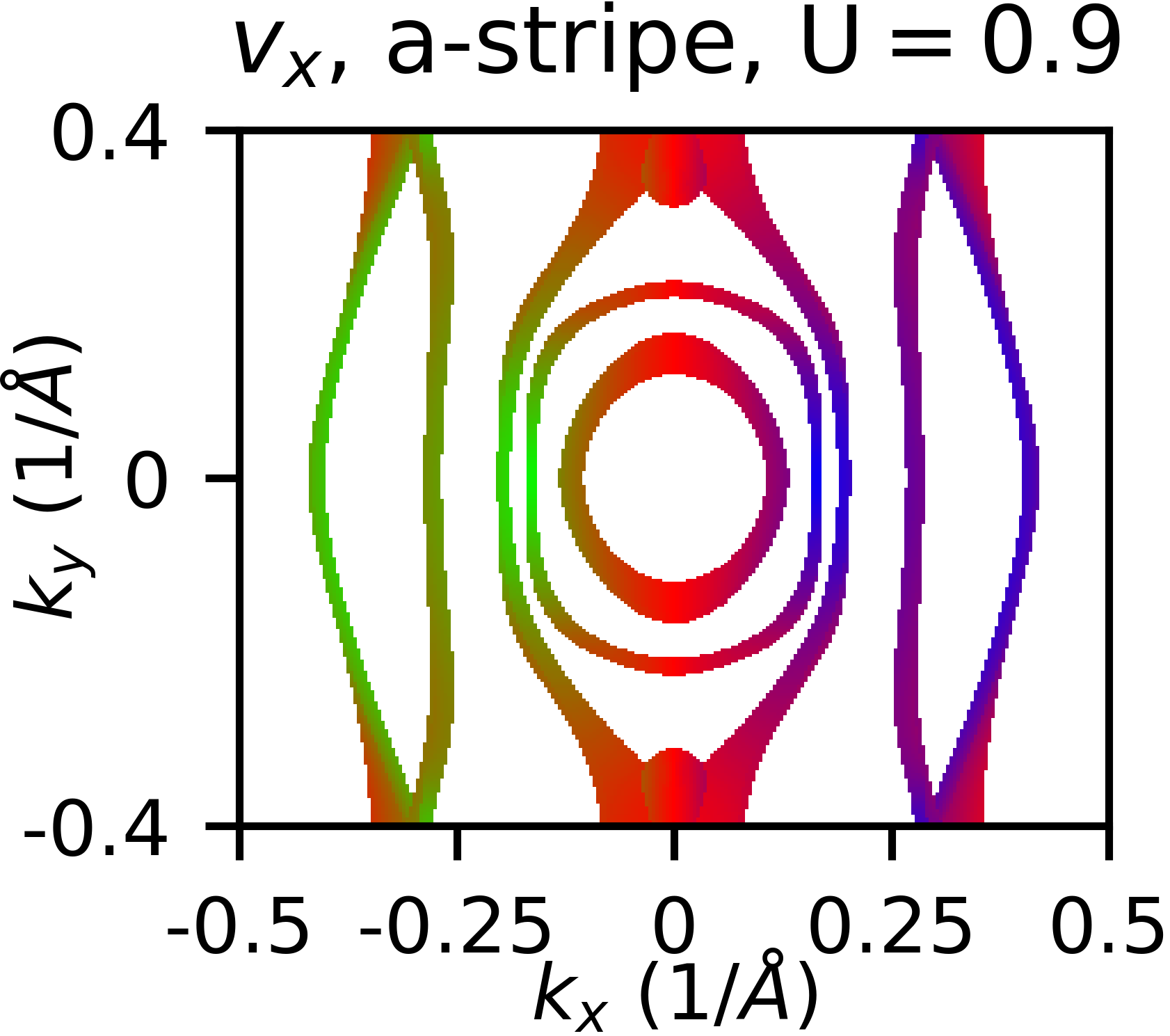}
\end{subfigure}
\begin{subfigure}[t]{0.48\columnwidth}\caption{}\label{fig:Vy_stripe_Up9}\includegraphics{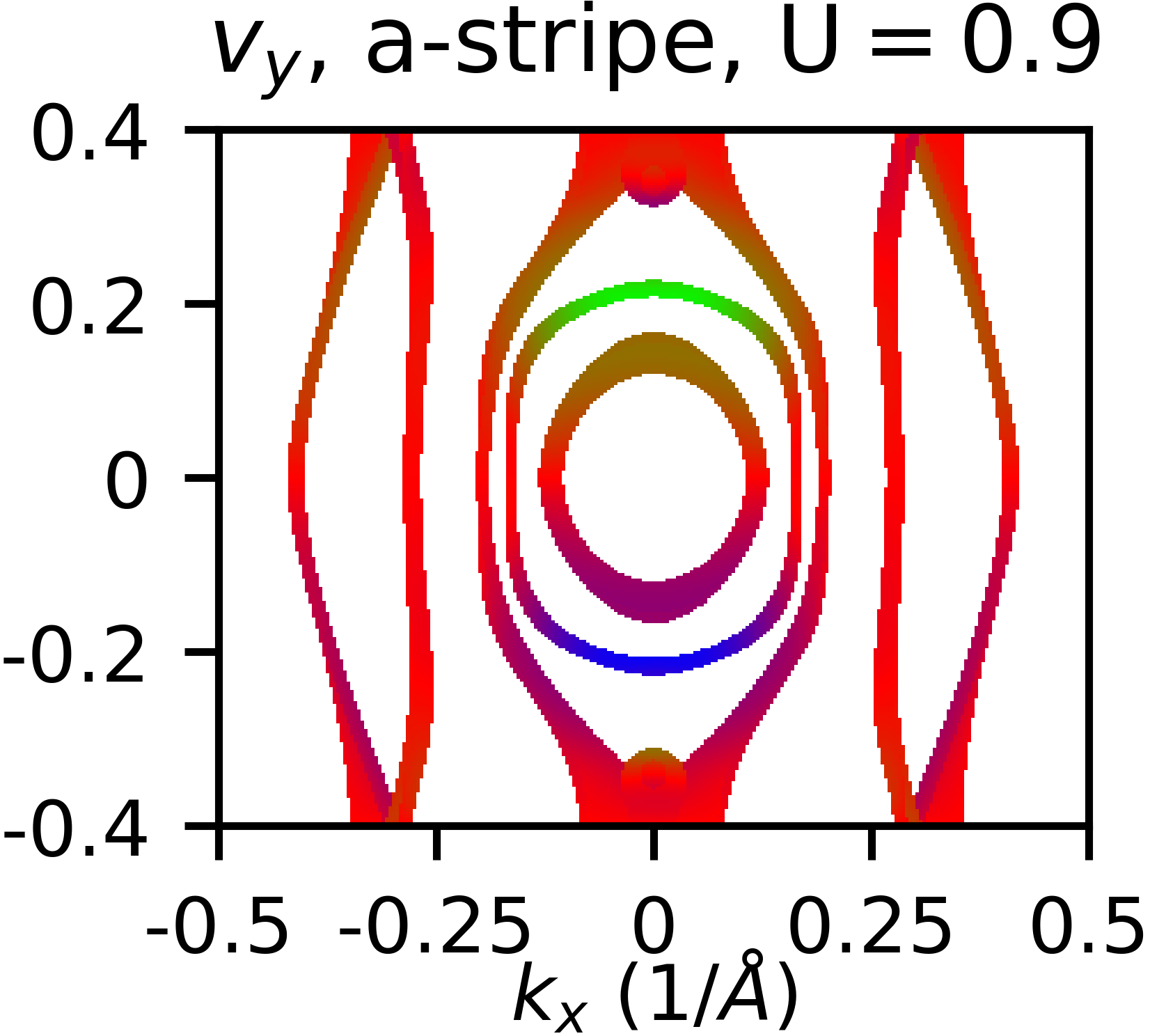}
\end{subfigure}
\begin{subfigure}[t]{0.48\columnwidth}\caption{}\label{fig:Vx_zig_Up9}\includegraphics{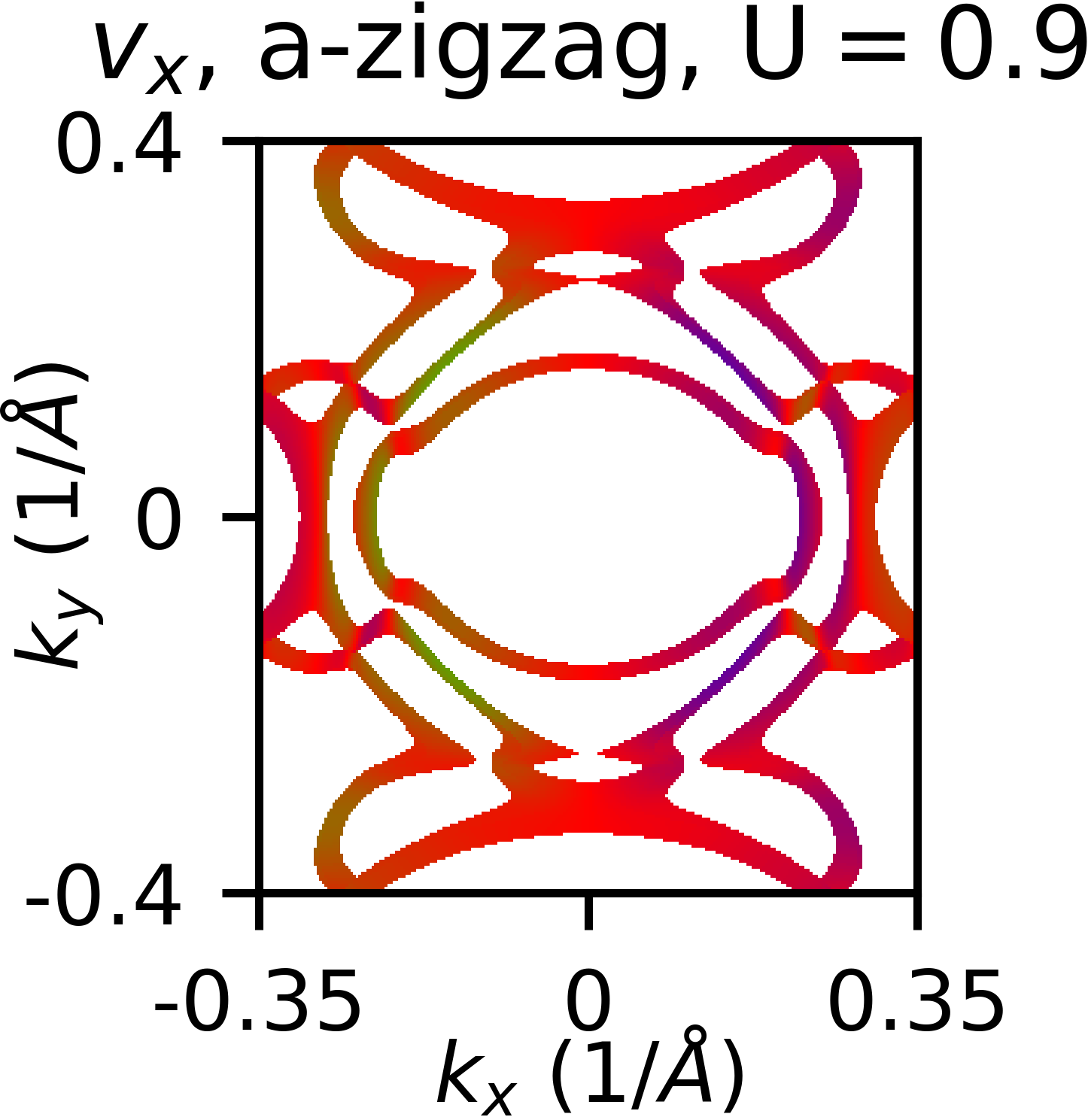}
\end{subfigure}
\begin{subfigure}[t]{0.48\columnwidth}\caption{}\label{fig:Vy_zig_Up9}\includegraphics{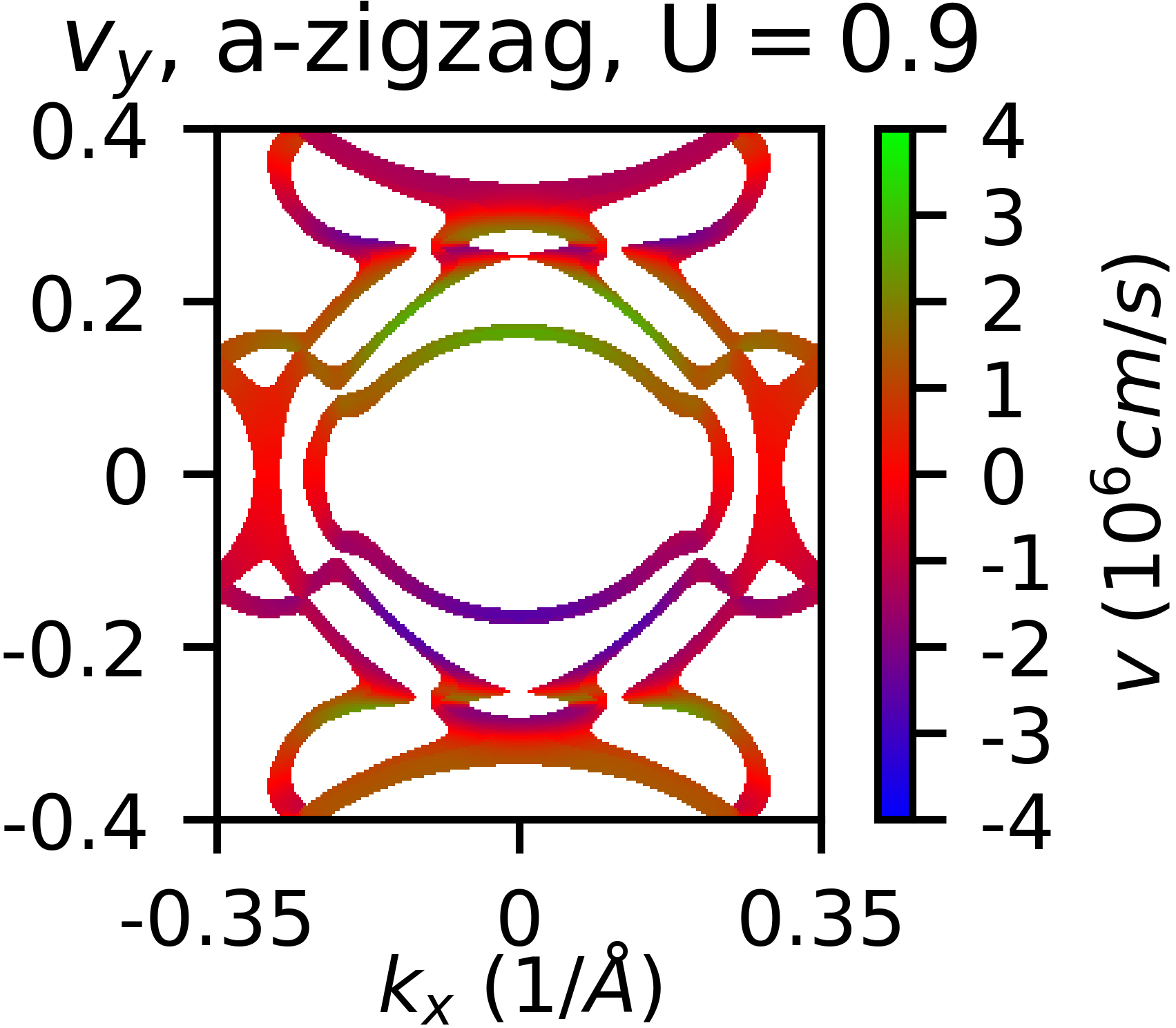}
\end{subfigure}
\caption{Electronic structure in the $k_z=0$ plane of $\mathrm{Fe_{1/3}NbS_2}$, with a finite broadening for aesthetic purposes of $10$ $\mathrm{meV}$ for a-stripe order with $\mathrm{U}=0.3$ ((a)-(b)) and $\mathrm{U}=0.9$ $\mathrm{eV}$ ((e)-(f)), and $2$ and $5$ $\mathrm{meV}$ for a-zigzag order with $\mathrm{U}=0.3$ ((c)-(d)) and $\mathrm{U}=0.9$ $\mathrm{eV}$ ((g)-(h)) respectively. The plots are colored by either the $\mathrm{x}$ or $\mathrm{y}$ component of band velocity, as indicated by the title.}\label{fig: FS} 
\end{figure*}

We now examine cross sections of the Fermi surfaces (FSs) for a-stripe and a-zigzag order computed with our two sets of PBE+U calculations. We focus on electronic structure parallel to the $k_x$-$k_y$ plane, relevant to the switching experiments. We plot Fermi contours in the $k_z=0$ plane of the Brillouin zone (BZ); cuts of the $k_x$-$k_y$ FS at other values of $k_z$ are given in the Supplement\cite{suppmat}. We focus first on the a-stripe FS, depicted in Figures \ref{fig:Vx_stripe_Up3}-\ref{fig:Vy_stripe_Up3} and \ref{fig:Vx_stripe_Up9}-\ref{fig:Vy_stripe_Up9} for both $\mathrm{U}=0.3$ and $\mathrm{U}=0.9$ $\mathrm{eV}$ respectively. We consider the two $\mathrm{U}$ values for the reasons discussed in Section \ref{sec:methods}. For both choices of $\mathrm{U}$, the a-stripe FS results from relatively flat bands extending along the entire $k_y$ direction of the BZ ($k_x$ is parallel to the $[100]$ crystallographic direction in real space, and $k_y$ parallel to $[120]$; we use the hexagonal notation of the primitive cell for crystallographic directions through the text.) We gain a more explicit picture of the corresponding anisotropy in carrier transport by examining the in-plane components of the band velocities. Figures \ref{fig:Vx_stripe_Up3} and \ref{fig:Vx_stripe_Up9} are color-coded according to $v_x(k_0)=\frac{1}{\hbar} \frac{ \partial E}{\partial k_x} |_{\mathbf{k}=k_0,E=E_F}$, where $x$ is along [$100]$, $E_F$ is the Fermi energy, and $k_0$ is a point in the $k_x$-$k_y$ plane. Figures \ref{fig:Vy_stripe_Up3} and \ref{fig:Vy_stripe_Up9} are colored by $v_y$, whose magnitude is greatly reduced compared to $v_x$. This suggests that, for the stripe phase, the conductance $\sigma_{xx}$ along the $x$ direction of the sample (parallel to the magnetic stripes in real space) will be higher than $\sigma_{yy}$ (perpendicular to the stripes ); and equivalently, the resistance $R_{xx}<R_{yy}$ for a-stripe order.\\
\indent While still anisotropic, the a-zigzag FS cuts, depicted in Figures \ref{fig:Vx_zig_Up3}-\ref{fig:Vy_zig_Up3} and \ref{fig:Vx_zig_Up9}-\ref{fig:Vy_zig_Up9} for $\mathrm{U}=0.3$ and $\mathrm{U}=0.9$ $\mathrm{eV}$, are more symmetric as compared to a-stripe. This is also evident from examining the band velocities. For PBE+U with $\mathrm{U}=0.3$ the $v_x$ and $v_y$ components at $E_F$ appear isotropic (Figures \ref{fig:Vx_zig_Up3} and \ref{fig:Vy_zig_Up3}), likely a coincidental result due to this choice of $\mathrm{U}$. The a-zigzag weight of $v_y$ relative to $v_x$ increases significantly for $\mathrm{U}=0.9$ $\mathrm{eV}$ (Figures \ref{fig:Vx_zig_Up9} and \ref{fig:Vy_zig_Up9}). This implies that that the transport anisotropy in a-zigzag, at least for $\mathrm{U}=0.9$ $\mathrm{eV}$, switches compared to stripe (i.e. for a-zigzag, $\sigma_{xx}<\sigma_{yy}$ and $R_{xx}>R_{yy}$). We point out that the large qualitative changes in the FS cross section for a-zigzag order in going from $\mathrm{U}=0.3$ to $\mathrm{U}=0.9$ $\mathrm{eV}$ as compared to a-stripe order are presumably linked to the large number of low-dispersion bands near the Fermi level for a-zigzag which are highly sensitive to small changes in $\mathrm{U}$ (see Supplement for orbital-projected band structures\cite{suppmat}).\\
\section{Resistivity tensor and Switching} 
\begin{figure}
\begin{subfigure}[t]{0.9\columnwidth}\caption{}\label{fig:device}\includegraphics[width=0.8\columnwidth]{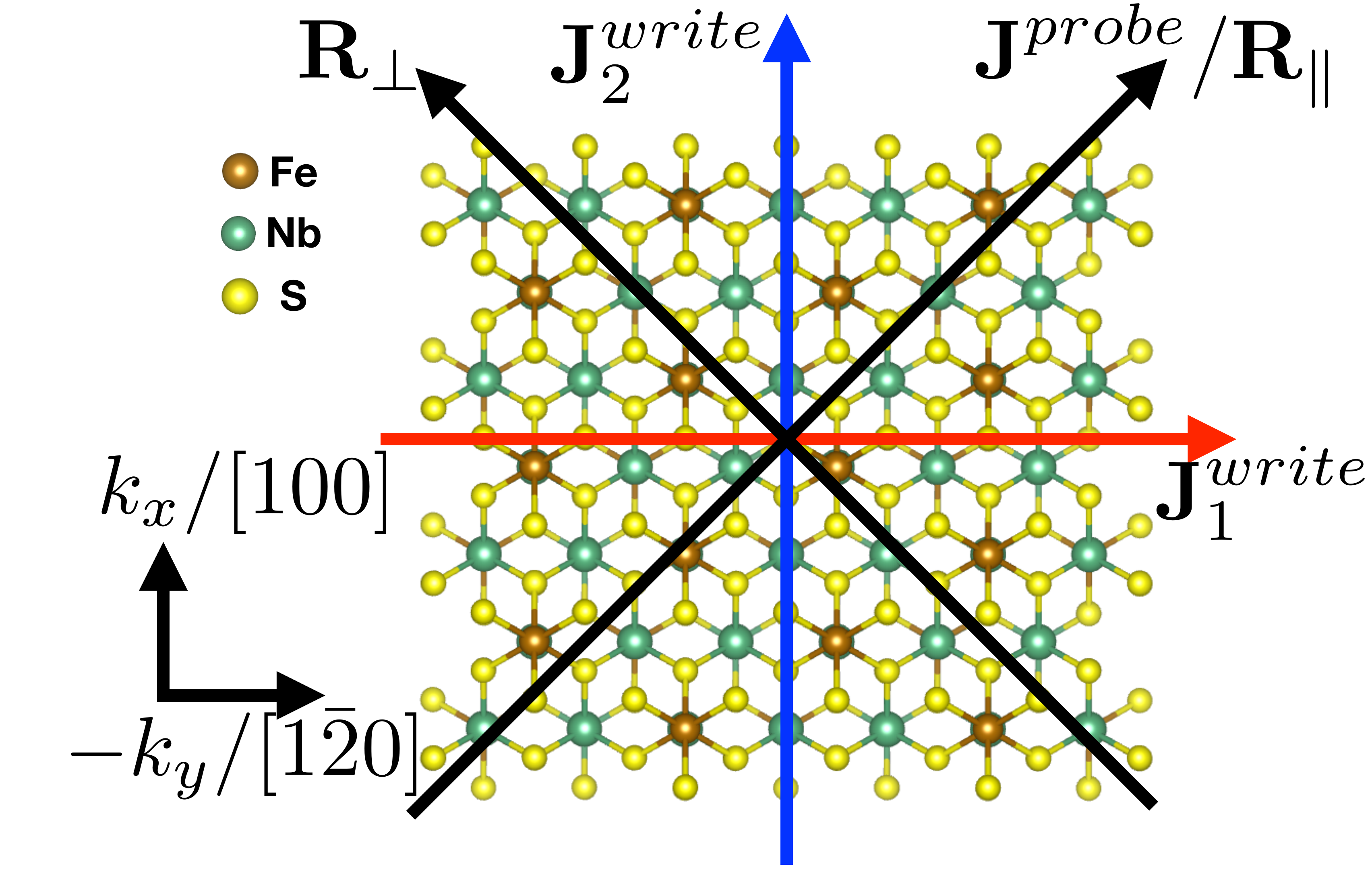}
\end{subfigure}
\begin{subfigure}[t]{0.52\columnwidth}\caption{}\label{fig:Rperp_stripe}\includegraphics{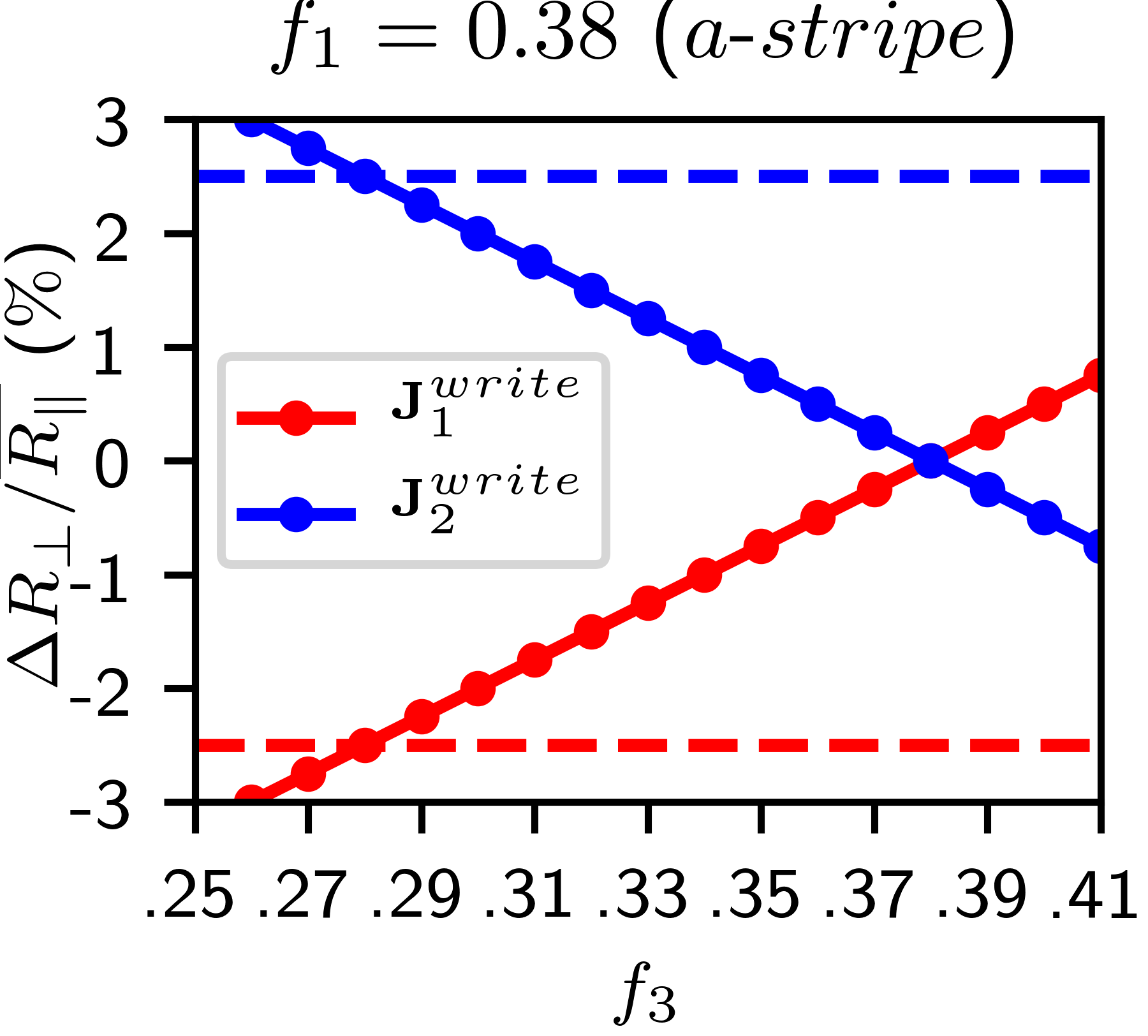}
\end{subfigure}
\begin{subfigure}[t]{0.42\columnwidth}\caption{}\label{fig:Rperp_zig}\includegraphics{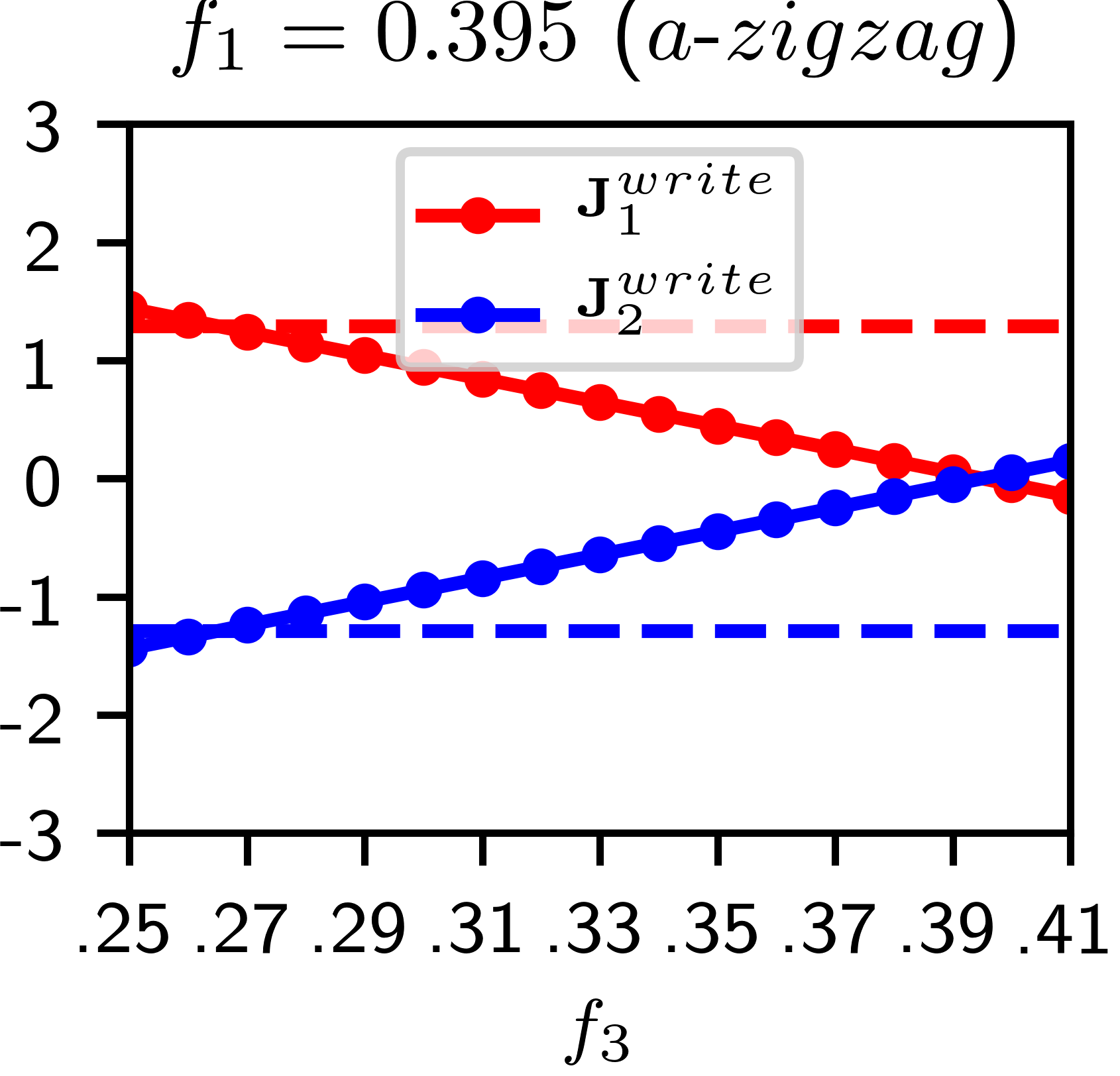}
\end{subfigure}
\caption{Electrical switching. (a) $\mathrm{Fe_{1/3}NbS_2}$ crystal structure overlaid with directions of applied currents and measured resistance in experiment. In the experiment, orthogonal pulses applied along the red and blue arrows switch $\mathrm{Fe_{1/3}NbS_2}$ between two states with different domain populations, detected by changes in the transverse resistance $\mathbf{R}_{\perp}$. (b)-(c) Calculated $\frac{\Delta \mathbf{R}_{\perp}}{\bar{\mathbf{R}_{\parallel}}}$ based on equations \ref{eq:domain_pulse1} (red) and \ref{eq:domain_pulse2} (blue) as a function of $f_3$ for a fixed initial value of $f_1$. $f_1$ ($f_3$) can be viewed as the resulting fractional population of the domain with principle axis along $[100]$ after $\mathbf{J}_1^{write}$ ($\mathbf{J}_2^{write}$). (b) corresponds to a-stripe phase, (c) corresponds to a-zigzag phase. Dashed lines (same color coding as the PBE+U-derived points) indicate the value of $f_3$ where the calculated $\frac{\Delta \mathbf{R}_{\perp}}{\bar{\mathbf{R}_{\parallel}}}$  agrees with the experimental data in Reference \citenum{Maniv2021} for $\mathrm{Fe}$ intercalations likely corresponding to a-stripe and a-zigzag order.}\label{fig: switching} 
\end{figure}

In order to understand the specific current-domain response implied by the FS anisotropies above, we can compute the resistivity tensor for mono-domain $\mathrm{Fe_{1/3}NbS_2}$ with input from our DFT calculations within the Kubo linear response formalism\cite{Mahan2000}. Within this formalism, using the eigenstate representation, the static conductivity tensor $\sigma$ in the zero-temperature limit may be written as \cite{Freimuth2014,Zelezny2017b}
\begin{multline}
\sigma_{ij}=-\frac{e\hbar}{\pi}\sum_{\mathbf{k},n,m}[\Gamma^2\Re(\bra{n\mathbf{k}}\hat{v}_i\ket{m\mathbf{k}} \bra{m\mathbf{k}}\hat{v}_j\ket{n\mathbf{k}})]\\
([(E_F-\epsilon_{n\mathbf{k}})^2+\Gamma^2] [(E_F-\epsilon_{m\mathbf{k}})^2+\Gamma^2])^{-1},
\label{eq:kubo}
\end{multline}
with $\epsilon_{n\mathbf{k}}$ the eigenenergy of the corresponding eigenstate $\ket{n\mathbf{k}}$ and $\hat{v}_i$ the velocity operator in the $ith$ direction. The indices $n$ and $m$ run over all bands (occupied and unoccupied). We use a constant band broadening $\Gamma$, where $\Gamma=\frac{\hbar}{2\tau}$ is inversely proportional to the electron relaxation time $\tau$, assuming $\tau$ is band and $\mathbf{k}$-independent, sufficient for our purposes. The Bloch eigenstates, eigenvalues, and velocity operators in Eq. \ref{eq:kubo} are constructed using Wannier functions obtained from our PBE+U calculations, and Equation \ref{eq:kubo} is evaluated using the Wannier Linear Response software\cite{Zelezny2017a}. In general, the linear-response conductivity can also contain a term which is odd under time reversal, whereas Equation \ref{eq:kubo} is even under this operation\cite{Zelezny2017b}. However, both a-stripe and a-zigzag magnetism possess time reversal symmetry plus a translation according to their magnetic space groups, such that the part of the conductivity which is odd under time reversal is necessarily zero, leaving us with only Equation \ref{eq:kubo} to evaluate.\\
\indent Apart from the approximations inherent in our conductivity tensors computed using Equation \ref{eq:kubo}, additional deviations from experimental results may come from our use of the pristine $x=\frac{1}{3}$ $\mathrm{Fe}$ concentration in all PBE+U calculations, as the recent transport and switching experiments\cite{Nair2019,Maniv2021} were performed on $\mathrm{Fe_xNbS_2}$ samples with a range of $\mathrm{Fe}$ concentrations $x\sim0.31-0.35$. Although NMR data suggests that a spin-glass coexists with the AFM order above and below $x=\frac{1}{3}$, and may well be the underlying mechanism for the efficient switching of the ordered magnetic domains\cite{Maniv2021}, we expect that the electronic structure and transport anisotropy of the stripe and zigzag phases, which we focus on in this paper, will not differ significantly between slightly off-stoichiometry structures and the $x=\frac{1}{3}$ structure we use in our DFT calculations. Moreover, the NMR measurements, as well as contemporary neutron experiments\cite{Wu2021}, find evidence for a slight in-plane magnetic moment in contrast to the earlier neutron studies\cite{VanLaar1971,Suzuki1993}. However, given the strong magnetic anisotropy which favors spins to point along the c axis in $\mathrm{Fe_{1/3}NbS_2}$\cite{Friend1977,Haley2020a}, we expect our focus on calculations of transport properties with collinear magnetic order along $\mathrm{c}$ to be an acceptable simplification.\\
 \indent Having obtained conductivity tensors within the constant relaxation time approximation, the resistance $\mathrm{R}$ is then the resistivity $\rho=\sigma^{-1}$ multiplied by the ratio of device length to cross-sectional area ($\sim3.7\times10^{-4}$ cm)\textsuperscript{-1}\cite{Nair2019}. In order to meaningfully compare the anisotropy of the resistance tensors with different magnetic orders and $\mathrm{U}$ values, we treat $\Gamma$ as a parameter and adjust it for each $\mathrm{U}$ and magnetic order such that the $R_{xx}$ component of the tensor (corresponding to the resistance along the [100] direction) is roughly equivalent to the experimentally measured resistance of $\mathrm{Fe_{1/3}NbS_2}$ samples, between $0.25$-$0.3$ $\Omega$\cite{communManiv2020}. Since the samples associated with these values are not mono-domain\cite{Nair2019}, this measured value does not, strictly speaking, correspond to the $R_{xx}$ of a single domain crystal, but we use it nonetheless to normalize the computed resistance tensors. We present the quantitative dependence of the resistance, as well as the in-plane anisotropy, on $\Gamma$ for each  magnetic ordering and $\mathrm{U}$ value in the Supplement\cite{suppmat}.\\ 
\indent The results of our calculations appear in Table \ref{tab:transport}. The transport anisotropy we compute from our PBE+U calculations, which we define quantitatively as $A=\frac{R_{yy}}{R_{xx}}$, is consistent with the calculated band velocities in Figure \ref{fig: FS}. For a-stripe ordering, $R_{yy}$ along $[120]$ is higher than $R_{xx}$ along [100] by roughly a factor of 2, for both $\mathrm{U}$ values considered. With a-zigzag ordering however, $R_{yy}$ becomes smaller than $R_{xx}$ ($A<1$). For both sets of PBE+U calculations, the transport anisotropy for a-zigzag is significantly reduced compared with stripe ordering. Indeed, for $\mathrm{U}=0.3$ $\mathrm{eV}$ the transport anisotropy is nearly unity for zigzag ordering.\\
\begin{table}
\caption{In-plane transport anisotropy computed for $\mathrm{Fe_{1/3}NbS_2}$, defined as $A=\frac{R_{yy}}{R_{xx}}$ with $\mathrm{x}$ along [100], for a-zigzag and a-stripe phases for both $\mathrm{U}$ values used in our PBE+U calculations. Absolute values of $R_{xx}$ and the values of $\Gamma$ used in Equation \ref{eq:kubo} are provided as well.}\label{tab:transport} 
	\begin{tabular}{| c | c | c | c | c |}
 \hline
&  \multicolumn{2}{|c|}{$U=0.3$ $\mathrm{eV}$} 
 & 
 \multicolumn{2}{|c|}{$U=0.9$ $\mathrm{eV}$} \\ \hline
& a-stripe & a-zigzag & a-stripe & a-zigzag\\
 \hline
$\Gamma$ ($\mathrm{meV}$) & $10$ & $5$ & $30$ & $10$\\
\hline
$R_{xx}$ ($\Omega$) & $0.26$ & 0.28 & 0.28 & 0.25\\
\hline
$A$ & 2.15 & 0.97 & 2.00 & 0.77\\
\hline
\end{tabular}
\end{table}
\indent Having obtained approximate resistivity tensors for mono-domain $\mathrm{Fe_{1/3}NbS_2}$ with a-stripe and a-zigzag ordering based on our PBE+U calculations, we can infer the current-domain response by comparing with experiment. In the following discussion we use our PBE+U results with $\mathrm{U}=0.9$ $\mathrm{eV}$. In Figure \ref{fig:device} we show the $\mathrm{a}$-$\mathrm{b}$ plane of the $\mathrm{Fe_{1/3}NbS_2}$ crystal overlaid with the directions of applied currents and measured resistances for the experiments in References \citenum{Nair2019} and \citenum{Maniv2021}. In these experiments, DC pulses, $\mathbf{J}_1^{write}$ and $\mathbf{J}_2^{write}$, were applied in succession along the $-k_y$/$[1\bar2{0}]$ and $k_x$/$[100]$ crystallographic directions. The low-frequency AC current $\mathbf{J}^{probe}$ used to measure the sample resistance after each writing pulse was applied at an angle of $45^{\circ}$ with respect to DC pulses. The transverse resistance $\mathbf{R}_{\perp}$ was read out along the contact which is orthogonal to $\mathbf{J}^{probe}$. Note that this is equal to the $\mathbf{R}_{xy}$ component of the resistance tensor with $x$ axis along $\mathbf{J}^{probe}$; we obtain this tensor by a rotation of our computed resistance matrix with $x$ axis along $[100]$\cite{Zhang2016} (see Supplementary material for details\cite{suppmat}).
The experimental changes in $\mathbf{R}_{\perp}$, normalized by the longitudinal resistance $\mathbf{R}_{\parallel}$ along $\mathbf{J}^{probe}$, are shown in Reference \citenum{Maniv2021} to be $\sim 2.5\%$ and $\sim1.3\%$ (when normalized to the same DC pulse current density) for $\mathrm{Fe}$ intercalations corresponding to $x=0.31$ and $x=0.35$ respectively; the smaller intercalation was used in Reference \citenum{Nair2019} as well. In addition to the reduction in magnitude of $\frac{\Delta \mathbf{R}_{\perp}}{\mathbf{R}_{\parallel}}$ going from the under-intercalated to over-intercalated sample, the sign of resistance change also switches; specifically, for $x=0.31$ a pulse along $\mathbf{J}_1^{write}$ causes a decrease in $\mathbf{R}_{\perp}$ whereas for $x=0.35$, $\Delta \mathbf{R}_{\perp}$ is positive after a pulse along $\mathbf{J}_1^{write}$. In interpreting the experimental results, we assume that $x=0.31$ and $x=0.35$ correspond to a-stripe and a-zigzag order respectively, as implied by neutron measurements (in addition to the results by Van Laar and Suzuki\cite{VanLaar1971,Suzuki1993}, a recent more systematic analysis of $\mathrm{Fe}$ concentration specifically indicates a stripe ground state for $x<\frac{1}{3}$ and a zigzag AFM ground state for $x>\frac{1}{3}$\cite{Wu2021}.) We note also that both zigzag and stripe magnetic space groups are consistent with the three-fold AFM domain structure observed by Little et al. (where the zigzags/stripe directions for each domain are related by $120^{\circ}$ rotations about $\mathrm{c}$\cite{Little2020}.)\\
\indent With these assumptions of the experimental magnetic order, we can explore the implications of our computed resistance tensors for domain repopulation with a-stripe and a-zigzag magnetism. We assume the total transverse resistance after each $\mathbf{J}_1^{write}$ or $\mathbf{J}_2^{write}$ pulse is proportional to the sum of resistances of the three domains, weighted by their fractional areas $f$, analogously to previous studies of domain-based anisotropic magnetoresistance\cite{Kriegner2016}. Then, we have
\begin{align}
[1\bar{2}0]/\mathbf{J}_1^{write}\rightarrow & \mathbf{R}_{\perp}^{\mathbf{J}_1^{write}}= f_1\mathbf{R}_{\perp}^{[100]}+f_2(\mathbf{R}_{\perp}^{[010]}+\mathbf{R}_{\perp}^{[\bar{1}\bar{1}0]}); \label{eq:domain_pulse1}
\end{align}
and
\begin{align}
[100]/\mathbf{J}_2^{write}\rightarrow & \mathbf{R}_{\perp}^{\mathbf{J}_2^{write}}=f_3\mathbf{R}_{\perp}^{[100]}+f_4(\mathbf{R}_{\perp}^{[010]}+\mathbf{R}_{\perp}^{[\bar{1}\bar{1}0]}),\label{eq:domain_pulse2}
\end{align}
where $\mathbf{R}_{\perp}^{[010]}$ for example is the transverse resistance for a single domain with principle axis along $[010]$. $f_1$ and $f_2$ are fractional domain populations after a $\mathbf{J}_1^{write}$ pulse, $f_3$ and $f_4$ result from a $\mathbf{J}_2^{write}$ pulse, and we set $f_2=(1-f_1)/2$ and $f_4=(1-f_3)/2$ in equations \ref{eq:domain_pulse1} and \ref{eq:domain_pulse2} to ensure the fractions add to unity. We assume in each case that $f([010])=f([\bar{1}\bar{1}0])$ because both writing pulses bisect these two axes; the resistance tensors for the three domains are connected by rotations of $2\pi/3$ (see Supplementary material\cite{suppmat}). The $\mathbf{R}_{\perp}$ values in equations \ref{eq:domain_pulse1} and \ref{eq:domain_pulse2} are obtained from the off-diagonal components of these tensors.\\
\indent We can calculate the relative fractional domain changes required to reproduce the experimental switching amplitudes for the pulses, defined as
\begin{equation}
\frac{\Delta \mathbf{R}_{\perp}^{\mathbf{J}_1^{write}/\mathbf{J}_2^{write}}}{\bar{\mathbf{R}}_{\parallel}}=\frac{ \mathbf{R}_{\perp}^{\mathbf{J}_1^{write}/\mathbf{J}_2^{write}}-\bar{\mathbf{R}}_{\perp}}{\bar{\mathbf{R}}_{\parallel}},
\label{eq:deltR}
\end{equation}
where
\begin{equation}
 \bar{\mathbf{R}}_{\perp}=(\mathbf{R}_{\perp}^{\mathbf{J}_1^{write}}+\mathbf{R}_{\perp}^{\mathbf{J}_2^{write}})/2; \bar{\mathbf{R}}_{\parallel}=(\mathbf{R}_{\parallel}^{\mathbf{J}_1^{write}}+\mathbf{R}_{\parallel}^{\mathbf{J}_2^{write}})/2
\label{eq:avg}
\end{equation}
are the averages of the two resistances. We do this by selecting constant values of $f_1$ (fraction of $[100]$ domain after $\mathbf{J}_1^{write}$) and plotting $\frac{\Delta \mathbf{R}_{\perp}}{\bar{\mathbf{R}}_{\parallel}}$ for both $\mathbf{J}_1^{write}$ and $\mathbf{J}_2^{write}$ as a function of $f_3$ (fraction of $[100]$ domain after $\mathbf{J}_2^{write}$). Note that $\Delta \mathbf{R}_{\perp}^{\mathbf{J}_1^{write}}/\bar{\mathbf{R}}_{\parallel}$ and $\Delta \mathbf{R}_{\perp}^{\mathbf{J}_2^{write}}/\bar{\mathbf{R}}_{\parallel}$ are each dependent on both $f_1$ and $f_3$ through $\bar{\mathbf{R}}_{\perp}$ and $\bar{\mathbf{R}}_{\parallel}$ defined in equation \ref{eq:avg}. Results based on our PBE+U (with $\mathrm{U}=0.9$ $\mathrm{eV}$) calculations are shown in Figures \ref{fig:Rperp_stripe} and \ref{fig:Rperp_zig}. In both plots we have selected $f_1$ such that the values of $f_1$ and $f_3$ which yield the experimental resistance changes are symmetrically displaced about $f=\frac{1}{3}$, which is the equilibrium fraction we would expect for all three domains in the absence of external current. We emphasize however that for a given magnetic order, the qualitative results are identical regardless of the value of $f_1$, i.e. the sign and magnitude of the fractional change $f_1-f_3$ of domain $[100]$ between the the pulses remains constant. The dashed lines correspond to the experimental percent values for the intercalation corresponding to the same magnetic order. We see that, as a consequence of the crossover in the computed anisotropy from $A>1$ for a-stripe to $A<1$ for a-zigzag, the current-domain response for both magnetic structures is the same assuming the experimental data with opposite $\Delta \mathbf{R}_{\perp}$ signs indeed corresponds to the two proposed magnetic orders. Specifically, to replicate the correct sign of switching from experiment, for \emph{both} a-stripe and a-zigzag order, $f_1-f_3>0$. This means that the $\mathbf{J}_1^{write}$ pulse along $[1\bar{2}0]$ causes a fractional increase in the orthogonal $[100]$ domain, whereas the $\mathbf{J}_2^{write}$ pulse parallel to $[100]$ destabilizes the $[100]$ domain and increases the fraction of domains alongs $[010]$ and $[\bar{1}\bar{1}0]$. Moreover, we can see that experimental reduction in switching amplitude for a-zigzag order compared to a-stripe is consistent with the reduced in-plane anisotropy we find for a-zigzag order in our PBE+U calculations. Indeed, using our $\mathrm{U}=0.9$ $\mathrm{eV}$ PBE+U results, the computed fractional changes from the equilibrium distribution $\frac{1}{3}:\frac{1}{3}:\frac{1}{3}$ required to match the corresponding experimental resistance changes are very close, $f_1=0.38$ for a-stripe and $f_1=0.395$ for a-zigzag, as one would expect for a given current density.\\
\section{Discussion and Conclusion}
We have used DFT calculations to understand the magnetism and origins of the electrical switching observed in $\mathrm{Fe_{1/3}NbS_2}$. Our PBE+U calculations indicate that the experimentally proposed a-stripe and a-zigzag magnetic phases are nearly degenerate, consistent with neutron data\cite{VanLaar1971,Suzuki1993,Wu2021} indicating that the ground state switches for small changes in $\mathrm{Fe}$ concentration. We find that the in-plane Fermi surface and corresponding transport for a-stripe order is anisotropic, with $R_{yy}>R_{xx}$, for all values of $\mathrm{U}$ used in our PBE+U calculations. The FS and transport for a-zigzag order is also anisotropic but the degree of anisotropy is reduced relative to stripe, and the quantitative results are highly sensitive to small changes in the Hubbard $\mathrm{U}$ used. Our findings suggest that there are two important factors leading to the particularly high anisotropy in electronic structure and transport for stripe order in $\mathrm{Fe_{1/3}NbS_2}$. Firstly, the reduction of six-fold symmetry in the high-temperature paramagnetic phase to two-fold symmetry due to the in-plane stripe magnetic order is consistent with the high anisotropy of the FS. Isostructural $\mathrm{Co_{1/3}NbS_2}$, also believed to have a stripe ground state, has been reported to have an anisotropic FS with quasi-flat bands much like $\mathrm{Fe_{1/3}NbS_2}$ from prior DFT calculations\cite{Popcevic2020}. With a-zigzag ordering however, the anisotropy in electronic structure and transport for $\mathrm{Fe_{1/3}NbS_2}$, while still present, is significantly reduced in spite of an identical reduction to two-fold rotational symmetry due to the magnetic order. This suggests that the magnetic interactions between nearest $\mathrm{Fe}$ neighbors may play an even larger role than rotational symmetry reduction in determining the degree of anisotropy in transport.\\
\indent Our calculations also reveal that, for both a-zigzag and a-stripe magnetic order, a pulse along a given direction should disfavor domains whose principle axes (and stripes/zigzags) are parallel to the pulse, and increase the populations of the other two domains. This directional dependence has implications for the microscopic details of the mechanism responsible for the current-induced domain repopulation. Further studies are required to understand the origins of the current-domain coupling which leads to domains parallel to the current pulse being disfavored, and whether this is consistent with the spin glass-mediated spin-orbit torque mechanism proposed in Reference \citenum{Maniv2021}.\\
\indent To be more concrete, we explicitly mention two possible future experimental outcomes for which our computed current-domain response will be particularly relevant. First, if further neutron scattering studies show unambiguously that the $\mathrm{Fe}$ spins in $\mathrm{Fe_{1/3}NbS_2}$ have zero in-plane component, the origin of current-induced switching must differ from traditional spin-orbit torque mechanisms, including the spin-glass mediated case proposed in Reference \citenum{Maniv2021}. This is because the in-plane directionality of the spin-orbit torque in the experimental geometry could not result in a switching between domains with the N\'{e}el vector fully along [001] for all three domains. In this situation, knowledge of the directionality of domain stabilization could inform the search for a novel switching mechanism. Alternatively, further studies expanding on Reference \citenum{Maniv2021} may definitively establish the direction in which polarized electrons in the coexisting spin glass are rotating a small in-plane component of the N\'{e}el vector in the ordered a-stripe and a-zigzag phases studied in this manuscript (i.e., away from or toward the current). This information, combined with our finding that a current pulse destabilizes domains with principle axes parallel to the pulse, will indicate the likely direction of the in-plane N\'{e}el vector component for a given domain. To be specific, if the current is found to rotate the in-plane component of the N\'{e}el vector away from the current pulse, our current-domain response findings indicate that the in-plane component is along the direction of the domain principle axis (parallel to the stripes or zigzags). However, if the current tends to align the in-plane N\'{e}el vector component parallel to the pulse, this suggests that the small in-plane moment is perpendicular to the direction of the domain principle axis.  Overall, our transport and electronic structure calculations support repopulation of magnetic domains being the underlying cause of electrical switching in $\mathrm{Fe_{1/3}NbS_2}$, and provide a platform for future studies.
\begin{acknowledgments}
The authors wish to thank E. Maniv and S. Wu for invaluable discussions regarding the experimental data. This work is supported by the Center for Novel Pathways to Quantum Coherence in Materials, an Energy Frontier Research Center funded by the US Department of Energy, Director, Office of Science, Office of Basic Energy Sciences under Contract No. DE-AC02-05CH11231. Computational resources provided by the Molecular Foundry through the US Department of Energy, Office of Basic Energy Sciences, and the National Energy Research Scientific Computing Center (NERSC), under the same contract number. Calculations were also performed on the Lawrencium cluster, operated by Lawrence Berkeley National Laboratory, and on Savio, operated by the University of California, Berkeley. 
\end{acknowledgments}

\FloatBarrier
\bibliography{FeTMD_paper.bib}

\begin{thebibliography}{42}%
\makeatletter
\providecommand \@ifxundefined [1]{%
 \@ifx{#1\undefined}
}%
\providecommand \@ifnum [1]{%
 \ifnum #1\expandafter \@firstoftwo
 \else \expandafter \@secondoftwo
 \fi
}%
\providecommand \@ifx [1]{%
 \ifx #1\expandafter \@firstoftwo
 \else \expandafter \@secondoftwo
 \fi
}%
\providecommand \natexlab [1]{#1}%
\providecommand \enquote  [1]{``#1''}%
\providecommand \bibnamefont  [1]{#1}%
\providecommand \bibfnamefont [1]{#1}%
\providecommand \citenamefont [1]{#1}%
\providecommand \href@noop [0]{\@secondoftwo}%
\providecommand \href [0]{\begingroup \@sanitize@url \@href}%
\providecommand \@href[1]{\@@startlink{#1}\@@href}%
\providecommand \@@href[1]{\endgroup#1\@@endlink}%
\providecommand \@sanitize@url [0]{\catcode `\\12\catcode `\$12\catcode
  `\&12\catcode `\#12\catcode `\^12\catcode `\_12\catcode `\%12\relax}%
\providecommand \@@startlink[1]{}%
\providecommand \@@endlink[0]{}%
\providecommand \url  [0]{\begingroup\@sanitize@url \@url }%
\providecommand \@url [1]{\endgroup\@href {#1}{\urlprefix }}%
\providecommand \urlprefix  [0]{URL }%
\providecommand \Eprint [0]{\href }%
\providecommand \doibase [0]{http://dx.doi.org/}%
\providecommand \selectlanguage [0]{\@gobble}%
\providecommand \bibinfo  [0]{\@secondoftwo}%
\providecommand \bibfield  [0]{\@secondoftwo}%
\providecommand \translation [1]{[#1]}%
\providecommand \BibitemOpen [0]{}%
\providecommand \bibitemStop [0]{}%
\providecommand \bibitemNoStop [0]{.\EOS\space}%
\providecommand \EOS [0]{\spacefactor3000\relax}%
\providecommand \BibitemShut  [1]{\csname bibitem#1\endcsname}%
\let\auto@bib@innerbib\@empty
\bibitem [{\citenamefont {Manchon}\ \emph {et~al.}(2019)\citenamefont
  {Manchon}, \citenamefont {{\v{Z}}elezn{\'{y}}}, \citenamefont {Miron},
  \citenamefont {Jungwirth}, \citenamefont {Sinova}, \citenamefont {Thiaville},
  \citenamefont {Garello},\ and\ \citenamefont {Gambardella}}]{Manchon2019}%
  \BibitemOpen
  \bibfield  {author} {\bibinfo {author} {\bibfnamefont {A.}~\bibnamefont
  {Manchon}}, \bibinfo {author} {\bibfnamefont {J.}~\bibnamefont
  {{\v{Z}}elezn{\'{y}}}}, \bibinfo {author} {\bibfnamefont {I.~M.}\
  \bibnamefont {Miron}}, \bibinfo {author} {\bibfnamefont {T.}~\bibnamefont
  {Jungwirth}}, \bibinfo {author} {\bibfnamefont {J.}~\bibnamefont {Sinova}},
  \bibinfo {author} {\bibfnamefont {A.}~\bibnamefont {Thiaville}}, \bibinfo
  {author} {\bibfnamefont {K.}~\bibnamefont {Garello}}, \ and\ \bibinfo
  {author} {\bibfnamefont {P.}~\bibnamefont {Gambardella}},\ }\href {\doibase
  10.1103/RevModPhys.91.035004} {\bibfield  {journal} {\bibinfo  {journal}
  {Reviews of Modern Physics}\ }\textbf {\bibinfo {volume} {91}} (\bibinfo
  {year} {2019}),\ 10.1103/RevModPhys.91.035004},\ \Eprint
  {http://arxiv.org/abs/1801.09636} {arXiv:1801.09636} \BibitemShut {NoStop}%
\bibitem [{\citenamefont {Manchon}\ and\ \citenamefont
  {Zhang}(2008)}]{Manchon2008}%
  \BibitemOpen
  \bibfield  {author} {\bibinfo {author} {\bibfnamefont {A.}~\bibnamefont
  {Manchon}}\ and\ \bibinfo {author} {\bibfnamefont {S.}~\bibnamefont
  {Zhang}},\ }\href {\doibase 10.1103/PhysRevB.78.212405} {\bibfield  {journal}
  {\bibinfo  {journal} {Physical Review B - Condensed Matter and Materials
  Physics}\ }\textbf {\bibinfo {volume} {78}},\ \bibinfo {pages} {1} (\bibinfo
  {year} {2008})}\BibitemShut {NoStop}%
\bibitem [{\citenamefont {Manchon}\ and\ \citenamefont
  {Zhang}(2009)}]{Manchon2009}%
  \BibitemOpen
  \bibfield  {author} {\bibinfo {author} {\bibfnamefont {A.}~\bibnamefont
  {Manchon}}\ and\ \bibinfo {author} {\bibfnamefont {S.}~\bibnamefont
  {Zhang}},\ }\href {\doibase 10.1103/PhysRevB.79.094422} {\bibfield  {journal}
  {\bibinfo  {journal} {Physical Review B - Condensed Matter and Materials
  Physics}\ }\textbf {\bibinfo {volume} {79}},\ \bibinfo {pages} {1} (\bibinfo
  {year} {2009})}\BibitemShut {NoStop}%
\bibitem [{\citenamefont {Bel'kov}\ and\ \citenamefont
  {Ganichev}(2008)}]{Belkov2008}%
  \BibitemOpen
  \bibfield  {author} {\bibinfo {author} {\bibfnamefont {V.~V.}\ \bibnamefont
  {Bel'kov}}\ and\ \bibinfo {author} {\bibfnamefont {S.~D.}\ \bibnamefont
  {Ganichev}},\ }\href {\doibase 10.1088/0268-1242/23/11/114003} {\bibfield
  {journal} {\bibinfo  {journal} {Semiconductor Science and Technology}\
  }\textbf {\bibinfo {volume} {23}} (\bibinfo {year} {2008}),\
  10.1088/0268-1242/23/11/114003},\ \Eprint {http://arxiv.org/abs/0803.0949}
  {arXiv:0803.0949} \BibitemShut {NoStop}%
\bibitem [{\citenamefont {Fukami}\ and\ \citenamefont
  {Ohno}(2017)}]{Fukami2017}%
  \BibitemOpen
  \bibfield  {author} {\bibinfo {author} {\bibfnamefont {S.}~\bibnamefont
  {Fukami}}\ and\ \bibinfo {author} {\bibfnamefont {H.}~\bibnamefont {Ohno}},\
  }\href {\doibase 10.7567/JJAP.56.0802A1} {\bibfield  {journal} {\bibinfo
  {journal} {Japanese Journal of Applied Physics}\ }\textbf {\bibinfo {volume}
  {56}} (\bibinfo {year} {2017}),\ 10.7567/JJAP.56.0802A1}\BibitemShut
  {NoStop}%
\bibitem [{\citenamefont {Sinova}\ \emph {et~al.}(2015)\citenamefont {Sinova},
  \citenamefont {Valenzuela}, \citenamefont {Wunderlich}, \citenamefont
  {Back},\ and\ \citenamefont {Jungwirth}}]{Sinova2015}%
  \BibitemOpen
  \bibfield  {author} {\bibinfo {author} {\bibfnamefont {J.}~\bibnamefont
  {Sinova}}, \bibinfo {author} {\bibfnamefont {S.~O.}\ \bibnamefont
  {Valenzuela}}, \bibinfo {author} {\bibfnamefont {J.}~\bibnamefont
  {Wunderlich}}, \bibinfo {author} {\bibfnamefont {C.~H.}\ \bibnamefont
  {Back}}, \ and\ \bibinfo {author} {\bibfnamefont {T.}~\bibnamefont
  {Jungwirth}},\ }\href {\doibase 10.1103/RevModPhys.87.1213} {\bibfield
  {journal} {\bibinfo  {journal} {Reviews of Modern Physics}\ }\textbf
  {\bibinfo {volume} {87}},\ \bibinfo {pages} {1213} (\bibinfo {year}
  {2015})}\BibitemShut {NoStop}%
\bibitem [{\citenamefont {{\v{Z}}elezn{\'{y}}}\ \emph
  {et~al.}(2017{\natexlab{a}})\citenamefont {{\v{Z}}elezn{\'{y}}},
  \citenamefont {Gao}, \citenamefont {Manchon}, \citenamefont {Freimuth},
  \citenamefont {Mokrousov}, \citenamefont {Zemen}, \citenamefont
  {Ma{\v{s}}ek}, \citenamefont {Sinova},\ and\ \citenamefont
  {Jungwirth}}]{Zelezny2017}%
  \BibitemOpen
  \bibfield  {author} {\bibinfo {author} {\bibfnamefont {J.}~\bibnamefont
  {{\v{Z}}elezn{\'{y}}}}, \bibinfo {author} {\bibfnamefont {H.}~\bibnamefont
  {Gao}}, \bibinfo {author} {\bibfnamefont {A.}~\bibnamefont {Manchon}},
  \bibinfo {author} {\bibfnamefont {F.}~\bibnamefont {Freimuth}}, \bibinfo
  {author} {\bibfnamefont {Y.}~\bibnamefont {Mokrousov}}, \bibinfo {author}
  {\bibfnamefont {J.}~\bibnamefont {Zemen}}, \bibinfo {author} {\bibfnamefont
  {J.}~\bibnamefont {Ma{\v{s}}ek}}, \bibinfo {author} {\bibfnamefont
  {J.}~\bibnamefont {Sinova}}, \ and\ \bibinfo {author} {\bibfnamefont
  {T.}~\bibnamefont {Jungwirth}},\ }\href {\doibase 10.1103/PhysRevB.95.014403}
  {\bibfield  {journal} {\bibinfo  {journal} {Physical Review B}\ }\textbf
  {\bibinfo {volume} {95}},\ \bibinfo {pages} {1} (\bibinfo {year}
  {2017}{\natexlab{a}})},\ \Eprint {http://arxiv.org/abs/1604.07590}
  {arXiv:1604.07590} \BibitemShut {NoStop}%
\bibitem [{\citenamefont {Olejn{\'{i}}k}\ \emph {et~al.}(2018)\citenamefont
  {Olejn{\'{i}}k}, \citenamefont {Seifert}, \citenamefont {Ka{\v{s}}par},
  \citenamefont {Nov{\'{a}}k}, \citenamefont {Wadley}, \citenamefont {Campion},
  \citenamefont {Baumgartner}, \citenamefont {Gambardella}, \citenamefont
  {Nemec}, \citenamefont {Wunderlich}, \citenamefont {Sinova}, \citenamefont
  {Ku{\v{z}}el}, \citenamefont {M{\"{u}}ller}, \citenamefont {Kampfrath},\ and\
  \citenamefont {Jungwirth}}]{Olejnik2018}%
  \BibitemOpen
  \bibfield  {author} {\bibinfo {author} {\bibfnamefont {K.}~\bibnamefont
  {Olejn{\'{i}}k}}, \bibinfo {author} {\bibfnamefont {T.}~\bibnamefont
  {Seifert}}, \bibinfo {author} {\bibfnamefont {Z.}~\bibnamefont
  {Ka{\v{s}}par}}, \bibinfo {author} {\bibfnamefont {V.}~\bibnamefont
  {Nov{\'{a}}k}}, \bibinfo {author} {\bibfnamefont {P.}~\bibnamefont {Wadley}},
  \bibinfo {author} {\bibfnamefont {R.~P.}\ \bibnamefont {Campion}}, \bibinfo
  {author} {\bibfnamefont {M.}~\bibnamefont {Baumgartner}}, \bibinfo {author}
  {\bibfnamefont {P.}~\bibnamefont {Gambardella}}, \bibinfo {author}
  {\bibfnamefont {P.}~\bibnamefont {Nemec}}, \bibinfo {author} {\bibfnamefont
  {J.}~\bibnamefont {Wunderlich}}, \bibinfo {author} {\bibfnamefont
  {J.}~\bibnamefont {Sinova}}, \bibinfo {author} {\bibfnamefont
  {P.}~\bibnamefont {Ku{\v{z}}el}}, \bibinfo {author} {\bibfnamefont
  {M.}~\bibnamefont {M{\"{u}}ller}}, \bibinfo {author} {\bibfnamefont
  {T.}~\bibnamefont {Kampfrath}}, \ and\ \bibinfo {author} {\bibfnamefont
  {T.}~\bibnamefont {Jungwirth}},\ }\href {\doibase 10.1126/sciadv.aar3566}
  {\bibfield  {journal} {\bibinfo  {journal} {Science Advances}\ }\textbf
  {\bibinfo {volume} {4}},\ \bibinfo {pages} {1} (\bibinfo {year} {2018})},\
  \Eprint {http://arxiv.org/abs/1711.08444} {arXiv:1711.08444} \BibitemShut
  {NoStop}%
\bibitem [{\citenamefont {Wadley}\ \emph {et~al.}(2016)\citenamefont {Wadley},
  \citenamefont {Howells}, \citenamefont {{\v{Z}}elezn{\'{y}}}, \citenamefont
  {Andrews}, \citenamefont {Hills}, \citenamefont {Campion}, \citenamefont
  {Nov{\'{a}}k}, \citenamefont {Olejn{\'{i}}k}, \citenamefont {Maccherozzi},
  \citenamefont {Dhesi}, \citenamefont {Martin}, \citenamefont {Wagner},
  \citenamefont {Wunderlich}, \citenamefont {Freimuth}, \citenamefont
  {Mokrousov}, \citenamefont {Kune{\v{s}}}, \citenamefont {Chauhan},
  \citenamefont {Grzybowski}, \citenamefont {Rushforth}, \citenamefont
  {Edmond}, \citenamefont {Gallagher},\ and\ \citenamefont
  {Jungwirth}}]{Wadley2016}%
  \BibitemOpen
  \bibfield  {author} {\bibinfo {author} {\bibfnamefont {P.}~\bibnamefont
  {Wadley}}, \bibinfo {author} {\bibfnamefont {B.}~\bibnamefont {Howells}},
  \bibinfo {author} {\bibfnamefont {J.}~\bibnamefont {{\v{Z}}elezn{\'{y}}}},
  \bibinfo {author} {\bibfnamefont {C.}~\bibnamefont {Andrews}}, \bibinfo
  {author} {\bibfnamefont {V.}~\bibnamefont {Hills}}, \bibinfo {author}
  {\bibfnamefont {R.~P.}\ \bibnamefont {Campion}}, \bibinfo {author}
  {\bibfnamefont {V.}~\bibnamefont {Nov{\'{a}}k}}, \bibinfo {author}
  {\bibfnamefont {K.}~\bibnamefont {Olejn{\'{i}}k}}, \bibinfo {author}
  {\bibfnamefont {F.}~\bibnamefont {Maccherozzi}}, \bibinfo {author}
  {\bibfnamefont {S.~S.}\ \bibnamefont {Dhesi}}, \bibinfo {author}
  {\bibfnamefont {S.~Y.}\ \bibnamefont {Martin}}, \bibinfo {author}
  {\bibfnamefont {T.}~\bibnamefont {Wagner}}, \bibinfo {author} {\bibfnamefont
  {J.}~\bibnamefont {Wunderlich}}, \bibinfo {author} {\bibfnamefont
  {F.}~\bibnamefont {Freimuth}}, \bibinfo {author} {\bibfnamefont
  {Y.}~\bibnamefont {Mokrousov}}, \bibinfo {author} {\bibfnamefont
  {J.}~\bibnamefont {Kune{\v{s}}}}, \bibinfo {author} {\bibfnamefont {J.~S.}\
  \bibnamefont {Chauhan}}, \bibinfo {author} {\bibfnamefont {M.~J.}\
  \bibnamefont {Grzybowski}}, \bibinfo {author} {\bibfnamefont {A.~W.}\
  \bibnamefont {Rushforth}}, \bibinfo {author} {\bibfnamefont {K.}~\bibnamefont
  {Edmond}}, \bibinfo {author} {\bibfnamefont {B.~L.}\ \bibnamefont
  {Gallagher}}, \ and\ \bibinfo {author} {\bibfnamefont {T.}~\bibnamefont
  {Jungwirth}},\ }\href {\doibase 10.1126/science.aab1031} {\bibfield
  {journal} {\bibinfo  {journal} {Science}\ }\textbf {\bibinfo {volume}
  {351}},\ \bibinfo {pages} {587} (\bibinfo {year} {2016})},\ \Eprint
  {http://arxiv.org/abs/1503.03765} {arXiv:1503.03765} \BibitemShut {NoStop}%
\bibitem [{\citenamefont {Bodnar}\ \emph {et~al.}(2018)\citenamefont {Bodnar},
  \citenamefont {{\v{S}}mejkal}, \citenamefont {Turek}, \citenamefont
  {Jungwirth}, \citenamefont {Gomonay}, \citenamefont {Sinova}, \citenamefont
  {Sapozhnik}, \citenamefont {Elmers}, \citenamefont {Kla{\"{u}}i},\ and\
  \citenamefont {Jourdan}}]{Bodnar2018}%
  \BibitemOpen
  \bibfield  {author} {\bibinfo {author} {\bibfnamefont {S.~Y.}\ \bibnamefont
  {Bodnar}}, \bibinfo {author} {\bibfnamefont {L.}~\bibnamefont
  {{\v{S}}mejkal}}, \bibinfo {author} {\bibfnamefont {I.}~\bibnamefont
  {Turek}}, \bibinfo {author} {\bibfnamefont {T.}~\bibnamefont {Jungwirth}},
  \bibinfo {author} {\bibfnamefont {O.}~\bibnamefont {Gomonay}}, \bibinfo
  {author} {\bibfnamefont {J.}~\bibnamefont {Sinova}}, \bibinfo {author}
  {\bibfnamefont {A.~A.}\ \bibnamefont {Sapozhnik}}, \bibinfo {author}
  {\bibfnamefont {H.~J.}\ \bibnamefont {Elmers}}, \bibinfo {author}
  {\bibfnamefont {M.}~\bibnamefont {Kla{\"{u}}i}}, \ and\ \bibinfo {author}
  {\bibfnamefont {M.}~\bibnamefont {Jourdan}},\ }\href {\doibase
  10.1038/s41467-017-02780-x} {\bibfield  {journal} {\bibinfo  {journal}
  {Nature Communications}\ }\textbf {\bibinfo {volume} {9}},\ \bibinfo {pages}
  {1} (\bibinfo {year} {2018})},\ \Eprint {http://arxiv.org/abs/1706.02482}
  {arXiv:1706.02482} \BibitemShut {NoStop}%
\bibitem [{\citenamefont {Moriyama}\ \emph {et~al.}(2018)\citenamefont
  {Moriyama}, \citenamefont {Oda}, \citenamefont {Ohkochi}, \citenamefont
  {Kimata},\ and\ \citenamefont {Ono}}]{Moriyama2018}%
  \BibitemOpen
  \bibfield  {author} {\bibinfo {author} {\bibfnamefont {T.}~\bibnamefont
  {Moriyama}}, \bibinfo {author} {\bibfnamefont {K.}~\bibnamefont {Oda}},
  \bibinfo {author} {\bibfnamefont {T.}~\bibnamefont {Ohkochi}}, \bibinfo
  {author} {\bibfnamefont {M.}~\bibnamefont {Kimata}}, \ and\ \bibinfo {author}
  {\bibfnamefont {T.}~\bibnamefont {Ono}},\ }\href {\doibase
  10.1038/s41598-018-32508-w} {\bibfield  {journal} {\bibinfo  {journal}
  {Scientific Reports}\ }\textbf {\bibinfo {volume} {8}},\ \bibinfo {pages} {1}
  (\bibinfo {year} {2018})},\ \Eprint {http://arxiv.org/abs/1708.07682}
  {arXiv:1708.07682} \BibitemShut {NoStop}%
\bibitem [{\citenamefont {Chen}\ \emph {et~al.}(2018)\citenamefont {Chen},
  \citenamefont {Zarzuela}, \citenamefont {Zhang}, \citenamefont {Song},
  \citenamefont {Zhou}, \citenamefont {Shi}, \citenamefont {Li}, \citenamefont
  {Zhou}, \citenamefont {Jiang}, \citenamefont {Pan},\ and\ \citenamefont
  {Tserkovnyak}}]{Chen2018}%
  \BibitemOpen
  \bibfield  {author} {\bibinfo {author} {\bibfnamefont {X.~Z.}\ \bibnamefont
  {Chen}}, \bibinfo {author} {\bibfnamefont {R.}~\bibnamefont {Zarzuela}},
  \bibinfo {author} {\bibfnamefont {J.}~\bibnamefont {Zhang}}, \bibinfo
  {author} {\bibfnamefont {C.}~\bibnamefont {Song}}, \bibinfo {author}
  {\bibfnamefont {X.~F.}\ \bibnamefont {Zhou}}, \bibinfo {author}
  {\bibfnamefont {G.~Y.}\ \bibnamefont {Shi}}, \bibinfo {author} {\bibfnamefont
  {F.}~\bibnamefont {Li}}, \bibinfo {author} {\bibfnamefont {H.~A.}\
  \bibnamefont {Zhou}}, \bibinfo {author} {\bibfnamefont {W.~J.}\ \bibnamefont
  {Jiang}}, \bibinfo {author} {\bibfnamefont {F.}~\bibnamefont {Pan}}, \ and\
  \bibinfo {author} {\bibfnamefont {Y.}~\bibnamefont {Tserkovnyak}},\ }\href
  {\doibase 10.1103/PhysRevLett.120.207204} {\bibfield  {journal} {\bibinfo
  {journal} {Physical Review Letters}\ }\textbf {\bibinfo {volume} {120}},\
  \bibinfo {pages} {1} (\bibinfo {year} {2018})},\ \Eprint
  {http://arxiv.org/abs/1804.05462} {arXiv:1804.05462} \BibitemShut {NoStop}%
\bibitem [{\citenamefont {Friend}\ \emph {et~al.}(1977)\citenamefont {Friend},
  \citenamefont {Beal},\ and\ \citenamefont {Yoffe}}]{Friend1977}%
  \BibitemOpen
  \bibfield  {author} {\bibinfo {author} {\bibfnamefont {R.~H.}\ \bibnamefont
  {Friend}}, \bibinfo {author} {\bibfnamefont {A.~R.}\ \bibnamefont {Beal}}, \
  and\ \bibinfo {author} {\bibfnamefont {A.~D.}\ \bibnamefont {Yoffe}},\ }\href
  {\doibase 10.1080/14786437708232952} {\bibfield  {journal} {\bibinfo
  {journal} {Philosophical Magazine}\ }\textbf {\bibinfo {volume} {35}},\
  \bibinfo {pages} {1269} (\bibinfo {year} {1977})}\BibitemShut {NoStop}%
\bibitem [{\citenamefont {{Van Laar}}\ \emph {et~al.}(1971)\citenamefont {{Van
  Laar}}, \citenamefont {Rietveld},\ and\ \citenamefont {Ijdo}}]{VanLaar1971}%
  \BibitemOpen
  \bibfield  {author} {\bibinfo {author} {\bibfnamefont {B.}~\bibnamefont {{Van
  Laar}}}, \bibinfo {author} {\bibfnamefont {H.~M.}\ \bibnamefont {Rietveld}},
  \ and\ \bibinfo {author} {\bibfnamefont {D.~J.}\ \bibnamefont {Ijdo}},\
  }\href {\doibase 10.1016/0022-4596(71)90019-3} {\bibfield  {journal}
  {\bibinfo  {journal} {Journal of Solid State Chemistry}\ }\textbf {\bibinfo
  {volume} {3}},\ \bibinfo {pages} {154} (\bibinfo {year} {1971})}\BibitemShut
  {NoStop}%
\bibitem [{\citenamefont {Nair}\ \emph {et~al.}(2019)\citenamefont {Nair},
  \citenamefont {Maniv}, \citenamefont {John}, \citenamefont {Doyle},
  \citenamefont {Orenstein},\ and\ \citenamefont {Analytis}}]{Nair2019}%
  \BibitemOpen
  \bibfield  {author} {\bibinfo {author} {\bibfnamefont {N.~L.}\ \bibnamefont
  {Nair}}, \bibinfo {author} {\bibfnamefont {E.}~\bibnamefont {Maniv}},
  \bibinfo {author} {\bibfnamefont {C.}~\bibnamefont {John}}, \bibinfo {author}
  {\bibfnamefont {S.}~\bibnamefont {Doyle}}, \bibinfo {author} {\bibfnamefont
  {J.}~\bibnamefont {Orenstein}}, \ and\ \bibinfo {author} {\bibfnamefont
  {J.~G.}\ \bibnamefont {Analytis}},\ }\href {\doibase
  10.1038/s41563-019-0518-x} {\bibfield  {journal} {\bibinfo  {journal} {Nature
  Materials}\ } (\bibinfo {year} {2019}),\ 10.1038/s41563-019-0518-x},\ \Eprint
  {http://arxiv.org/abs/1907.11698} {arXiv:1907.11698} \BibitemShut {NoStop}%
\bibitem [{\citenamefont {Little}\ \emph {et~al.}(2020)\citenamefont {Little},
  \citenamefont {Lee}, \citenamefont {John}, \citenamefont {Doyle},
  \citenamefont {Maniv}, \citenamefont {Nair}, \citenamefont {Chen},
  \citenamefont {Rees}, \citenamefont {Venderbos}, \citenamefont {Fernandes},
  \citenamefont {Analytis},\ and\ \citenamefont {Orenstein}}]{Little2020}%
  \BibitemOpen
  \bibfield  {author} {\bibinfo {author} {\bibfnamefont {A.}~\bibnamefont
  {Little}}, \bibinfo {author} {\bibfnamefont {C.}~\bibnamefont {Lee}},
  \bibinfo {author} {\bibfnamefont {C.}~\bibnamefont {John}}, \bibinfo {author}
  {\bibfnamefont {S.}~\bibnamefont {Doyle}}, \bibinfo {author} {\bibfnamefont
  {E.}~\bibnamefont {Maniv}}, \bibinfo {author} {\bibfnamefont {N.~L.}\
  \bibnamefont {Nair}}, \bibinfo {author} {\bibfnamefont {W.}~\bibnamefont
  {Chen}}, \bibinfo {author} {\bibfnamefont {D.}~\bibnamefont {Rees}}, \bibinfo
  {author} {\bibfnamefont {J.~W.}\ \bibnamefont {Venderbos}}, \bibinfo {author}
  {\bibfnamefont {R.~M.}\ \bibnamefont {Fernandes}}, \bibinfo {author}
  {\bibfnamefont {J.~G.}\ \bibnamefont {Analytis}}, \ and\ \bibinfo {author}
  {\bibfnamefont {J.}~\bibnamefont {Orenstein}},\ }\href {\doibase
  10.1038/s41563-020-0681-0} {\bibfield  {journal} {\bibinfo  {journal} {Nature
  Materials}\ } (\bibinfo {year} {2020}),\
  10.1038/s41563-020-0681-0}\BibitemShut {NoStop}%
\bibitem [{\citenamefont {Grzybowski}\ \emph {et~al.}(2017)\citenamefont
  {Grzybowski}, \citenamefont {Wadley}, \citenamefont {Edmonds}, \citenamefont
  {Beardsley}, \citenamefont {Hills}, \citenamefont {Campion}, \citenamefont
  {Gallagher}, \citenamefont {Chauhan}, \citenamefont {Novak}, \citenamefont
  {Jungwirth}, \citenamefont {Maccherozzi},\ and\ \citenamefont
  {Dhesi}}]{Grzybowski2017}%
  \BibitemOpen
  \bibfield  {author} {\bibinfo {author} {\bibfnamefont {M.~J.}\ \bibnamefont
  {Grzybowski}}, \bibinfo {author} {\bibfnamefont {P.}~\bibnamefont {Wadley}},
  \bibinfo {author} {\bibfnamefont {K.~W.}\ \bibnamefont {Edmonds}}, \bibinfo
  {author} {\bibfnamefont {R.}~\bibnamefont {Beardsley}}, \bibinfo {author}
  {\bibfnamefont {V.}~\bibnamefont {Hills}}, \bibinfo {author} {\bibfnamefont
  {R.~P.}\ \bibnamefont {Campion}}, \bibinfo {author} {\bibfnamefont {B.~L.}\
  \bibnamefont {Gallagher}}, \bibinfo {author} {\bibfnamefont {J.~S.}\
  \bibnamefont {Chauhan}}, \bibinfo {author} {\bibfnamefont {V.}~\bibnamefont
  {Novak}}, \bibinfo {author} {\bibfnamefont {T.}~\bibnamefont {Jungwirth}},
  \bibinfo {author} {\bibfnamefont {F.}~\bibnamefont {Maccherozzi}}, \ and\
  \bibinfo {author} {\bibfnamefont {S.~S.}\ \bibnamefont {Dhesi}},\ }\href
  {\doibase 10.1103/PhysRevLett.118.057701} {\bibfield  {journal} {\bibinfo
  {journal} {Physical Review Letters}\ }\textbf {\bibinfo {volume} {118}},\
  \bibinfo {pages} {1} (\bibinfo {year} {2017})},\ \Eprint
  {http://arxiv.org/abs/1607.08478} {arXiv:1607.08478} \BibitemShut {NoStop}%
\bibitem [{\citenamefont {Kresse}\ and\ \citenamefont
  {Furthm{\"{u}}ller}(1996)}]{Kresse1996}%
  \BibitemOpen
  \bibfield  {author} {\bibinfo {author} {\bibfnamefont {G.}~\bibnamefont
  {Kresse}}\ and\ \bibinfo {author} {\bibfnamefont {J.}~\bibnamefont
  {Furthm{\"{u}}ller}},\ }\href {\doibase 10.1103/PhysRevB.54.11169} {\bibfield
   {journal} {\bibinfo  {journal} {Physical Review B}\ }\textbf {\bibinfo
  {volume} {54}},\ \bibinfo {pages} {11169} (\bibinfo {year} {1996})},\ \Eprint
  {http://arxiv.org/abs/0927-0256(96)00008} {arXiv:0927-0256(96)00008
  [10.1016]} \BibitemShut {NoStop}%
\bibitem [{\citenamefont {Perdew}\ \emph {et~al.}(1996)\citenamefont {Perdew},
  \citenamefont {Burke},\ and\ \citenamefont {Ernzerhof}}]{Perdew1996}%
  \BibitemOpen
  \bibfield  {author} {\bibinfo {author} {\bibfnamefont {J.~P.}\ \bibnamefont
  {Perdew}}, \bibinfo {author} {\bibfnamefont {K.}~\bibnamefont {Burke}}, \
  and\ \bibinfo {author} {\bibfnamefont {M.}~\bibnamefont {Ernzerhof}},\ }\href
  {\doibase 10.1103/PhysRevLett.77.3865} {\bibfield  {journal} {\bibinfo
  {journal} {Physical Review Letters}\ }\textbf {\bibinfo {volume} {77}},\
  \bibinfo {pages} {3865} (\bibinfo {year} {1996})},\ \Eprint
  {http://arxiv.org/abs/0927-0256(96)00008} {arXiv:0927-0256(96)00008
  [10.1016]} \BibitemShut {NoStop}%
\bibitem [{\citenamefont {Bl{\"{o}}chl}(1994)}]{Blochl1994}%
  \BibitemOpen
  \bibfield  {author} {\bibinfo {author} {\bibfnamefont {P.~E.}\ \bibnamefont
  {Bl{\"{o}}chl}},\ }\href {\doibase 10.1103/PhysRevB.50.17953} {\bibfield
  {journal} {\bibinfo  {journal} {Physical Review B}\ }\textbf {\bibinfo
  {volume} {50}},\ \bibinfo {pages} {17953} (\bibinfo {year} {1994})},\ \Eprint
  {http://arxiv.org/abs/arXiv:1408.4701v2} {arXiv:arXiv:1408.4701v2}
  \BibitemShut {NoStop}%
\bibitem [{\citenamefont {Bl{\"{o}}chl}\ \emph {et~al.}(1994)\citenamefont
  {Bl{\"{o}}chl}, \citenamefont {Jepsen},\ and\ \citenamefont
  {Andersen}}]{Blochl1994a}%
  \BibitemOpen
  \bibfield  {author} {\bibinfo {author} {\bibfnamefont {P.~E.}\ \bibnamefont
  {Bl{\"{o}}chl}}, \bibinfo {author} {\bibfnamefont {O.}~\bibnamefont
  {Jepsen}}, \ and\ \bibinfo {author} {\bibfnamefont {O.~K.}\ \bibnamefont
  {Andersen}},\ }\href@noop {} {\bibfield  {journal} {\bibinfo  {journal}
  {Physical Review B}\ }\textbf {\bibinfo {volume} {49}},\ \bibinfo {pages}
  {16223} (\bibinfo {year} {1994})}\BibitemShut {NoStop}%
\bibitem [{\citenamefont {Marzari}\ and\ \citenamefont
  {Vanderbilt}(1997)}]{Marzari1997}%
  \BibitemOpen
  \bibfield  {author} {\bibinfo {author} {\bibfnamefont {N.}~\bibnamefont
  {Marzari}}\ and\ \bibinfo {author} {\bibfnamefont {D.}~\bibnamefont
  {Vanderbilt}},\ }\href {\doibase 10.1103/PhysRevB.56.12847} {\bibfield
  {journal} {\bibinfo  {journal} {Physical Review B}\ }\textbf {\bibinfo
  {volume} {56}},\ \bibinfo {pages} {22} (\bibinfo {year} {1997})},\ \Eprint
  {http://arxiv.org/abs/9707145} {arXiv:9707145 [cond-mat]} \BibitemShut
  {NoStop}%
\bibitem [{\citenamefont {Mostofi}\ \emph {et~al.}(2014)\citenamefont
  {Mostofi}, \citenamefont {Yates}, \citenamefont {Pizzi}, \citenamefont {Lee},
  \citenamefont {Souza}, \citenamefont {Vanderbilt},\ and\ \citenamefont
  {Marzari}}]{Mostofi2014}%
  \BibitemOpen
  \bibfield  {author} {\bibinfo {author} {\bibfnamefont {A.~A.}\ \bibnamefont
  {Mostofi}}, \bibinfo {author} {\bibfnamefont {J.~R.}\ \bibnamefont {Yates}},
  \bibinfo {author} {\bibfnamefont {G.}~\bibnamefont {Pizzi}}, \bibinfo
  {author} {\bibfnamefont {Y.~S.}\ \bibnamefont {Lee}}, \bibinfo {author}
  {\bibfnamefont {I.}~\bibnamefont {Souza}}, \bibinfo {author} {\bibfnamefont
  {D.}~\bibnamefont {Vanderbilt}}, \ and\ \bibinfo {author} {\bibfnamefont
  {N.}~\bibnamefont {Marzari}},\ }\href {\doibase 10.1016/j.cpc.2014.05.003}
  {\bibfield  {journal} {\bibinfo  {journal} {Computer Physics Communications}\
  }\textbf {\bibinfo {volume} {185}},\ \bibinfo {pages} {2309} (\bibinfo {year}
  {2014})},\ \Eprint {http://arxiv.org/abs/0708.0650} {arXiv:0708.0650}
  \BibitemShut {NoStop}%
\bibitem [{\citenamefont {Yates}\ \emph {et~al.}(2007)\citenamefont {Yates},
  \citenamefont {Wang}, \citenamefont {Vanderbilt},\ and\ \citenamefont
  {Souza}}]{Yates2007}%
  \BibitemOpen
  \bibfield  {author} {\bibinfo {author} {\bibfnamefont {J.~R.}\ \bibnamefont
  {Yates}}, \bibinfo {author} {\bibfnamefont {X.}~\bibnamefont {Wang}},
  \bibinfo {author} {\bibfnamefont {D.}~\bibnamefont {Vanderbilt}}, \ and\
  \bibinfo {author} {\bibfnamefont {I.}~\bibnamefont {Souza}},\ }\href
  {\doibase 10.1103/PhysRevB.75.195121} {\bibfield  {journal} {\bibinfo
  {journal} {Physical Review B - Condensed Matter and Materials Physics}\
  }\textbf {\bibinfo {volume} {75}},\ \bibinfo {pages} {1} (\bibinfo {year}
  {2007})},\ \Eprint {http://arxiv.org/abs/0702554} {arXiv:0702554 [cond-mat]}
  \BibitemShut {NoStop}%
\bibitem [{sup()}]{suppmat}%
  \BibitemOpen
  \href@noop {} {\bibinfo  {journal} {See Supplemental Material at [URL to be
  inserted] for PBE+U details, energetics, band structures, Gamma-dependence on
  transport anisotropy, additional FS cuts, transport for U=4, and rotation of
  coordinate system for domains}\ }\BibitemShut {NoStop}%
\bibitem [{\citenamefont {Wu}\ \emph {et~al.}(2018)\citenamefont {Wu},
  \citenamefont {Zhang}, \citenamefont {Song}, \citenamefont {Troyer},\ and\
  \citenamefont {Soluyanov}}]{Wu2018}%
  \BibitemOpen
\bibfield  {journal} {  }\bibfield  {author} {\bibinfo {author} {\bibfnamefont
  {Q.~S.}\ \bibnamefont {Wu}}, \bibinfo {author} {\bibfnamefont {S.~N.}\
  \bibnamefont {Zhang}}, \bibinfo {author} {\bibfnamefont {H.~F.}\ \bibnamefont
  {Song}}, \bibinfo {author} {\bibfnamefont {M.}~\bibnamefont {Troyer}}, \ and\
  \bibinfo {author} {\bibfnamefont {A.~A.}\ \bibnamefont {Soluyanov}},\ }\href
  {\doibase 10.1016/j.cpc.2017.09.033} {\bibfield  {journal} {\bibinfo
  {journal} {Computer Physics Communications}\ }\textbf {\bibinfo {volume}
  {224}},\ \bibinfo {pages} {405} (\bibinfo {year} {2018})}\BibitemShut
  {NoStop}%
\bibitem [{\citenamefont {Zelezny}(2018)}]{Zelezny2018}%
  \BibitemOpen
  \bibfield  {author} {\bibinfo {author} {\bibfnamefont {J.}~\bibnamefont
  {Zelezny}},\ }\href@noop {} {\enquote {\bibinfo {title} {{Wannier Linear
  Response}},}\ } (\bibinfo {year} {2018})\BibitemShut {NoStop}%
\bibitem [{\citenamefont {Anisimov}\ \emph {et~al.}(1997)\citenamefont
  {Anisimov}, \citenamefont {Aryasetiawan},\ and\ \citenamefont
  {Lichtenstein}}]{Anisimov1997}%
  \BibitemOpen
  \bibfield  {author} {\bibinfo {author} {\bibfnamefont {V.~I.}\ \bibnamefont
  {Anisimov}}, \bibinfo {author} {\bibfnamefont {F.}~\bibnamefont
  {Aryasetiawan}}, \ and\ \bibinfo {author} {\bibfnamefont {A.~I.}\
  \bibnamefont {Lichtenstein}},\ }\href {\doibase 10.1088/0953-8984/9/4/002}
  {\bibfield  {journal} {\bibinfo  {journal} {Journal of Physics Condensed
  Matter}\ }\textbf {\bibinfo {volume} {9}},\ \bibinfo {pages} {767} (\bibinfo
  {year} {1997})}\BibitemShut {NoStop}%
\bibitem [{\citenamefont {Dudarev}\ and\ \citenamefont
  {Botton}(1998)}]{Dudarev1998}%
  \BibitemOpen
  \bibfield  {author} {\bibinfo {author} {\bibfnamefont {S.}~\bibnamefont
  {Dudarev}}\ and\ \bibinfo {author} {\bibfnamefont {G.}~\bibnamefont
  {Botton}},\ }\href {\doibase 10.1103/PhysRevB.57.1505} {\bibfield  {journal}
  {\bibinfo  {journal} {Physical Review B - Condensed Matter and Materials
  Physics}\ }\textbf {\bibinfo {volume} {57}},\ \bibinfo {pages} {1505}
  (\bibinfo {year} {1998})},\ \Eprint {http://arxiv.org/abs/0927-0256(96)00008}
  {arXiv:0927-0256(96)00008 [10.1016]} \BibitemShut {NoStop}%
\bibitem [{\citenamefont {Haley}\ \emph {et~al.}(2020)\citenamefont {Haley},
  \citenamefont {Weber}, \citenamefont {Cookmeyer}, \citenamefont {Parker},
  \citenamefont {Maniv}, \citenamefont {Maksimovic}, \citenamefont {John},
  \citenamefont {Doyle}, \citenamefont {Maniv}, \citenamefont {Ramakrishna},
  \citenamefont {Reyes}, \citenamefont {Singleton}, \citenamefont {Moore},
  \citenamefont {Neaton},\ and\ \citenamefont {Analytis}}]{Haley2020a}%
  \BibitemOpen
  \bibfield  {author} {\bibinfo {author} {\bibfnamefont {S.~C.}\ \bibnamefont
  {Haley}}, \bibinfo {author} {\bibfnamefont {S.~F.}\ \bibnamefont {Weber}},
  \bibinfo {author} {\bibfnamefont {T.}~\bibnamefont {Cookmeyer}}, \bibinfo
  {author} {\bibfnamefont {D.~E.}\ \bibnamefont {Parker}}, \bibinfo {author}
  {\bibfnamefont {E.}~\bibnamefont {Maniv}}, \bibinfo {author} {\bibfnamefont
  {N.}~\bibnamefont {Maksimovic}}, \bibinfo {author} {\bibfnamefont
  {C.}~\bibnamefont {John}}, \bibinfo {author} {\bibfnamefont {S.}~\bibnamefont
  {Doyle}}, \bibinfo {author} {\bibfnamefont {A.}~\bibnamefont {Maniv}},
  \bibinfo {author} {\bibfnamefont {S.~K.}\ \bibnamefont {Ramakrishna}},
  \bibinfo {author} {\bibfnamefont {A.~P.}\ \bibnamefont {Reyes}}, \bibinfo
  {author} {\bibfnamefont {J.}~\bibnamefont {Singleton}}, \bibinfo {author}
  {\bibfnamefont {J.~E.}\ \bibnamefont {Moore}}, \bibinfo {author}
  {\bibfnamefont {J.~B.}\ \bibnamefont {Neaton}}, \ and\ \bibinfo {author}
  {\bibfnamefont {J.~G.}\ \bibnamefont {Analytis}},\ }\href {\doibase
  10.1103/PhysRevResearch.2.043020} {\bibfield  {journal} {\bibinfo  {journal}
  {Physical Review Research}\ }\textbf {\bibinfo {volume} {2}},\ \bibinfo
  {pages} {1} (\bibinfo {year} {2020})},\ \Eprint
  {http://arxiv.org/abs/2002.02960} {arXiv:2002.02960} \BibitemShut {NoStop}%
\bibitem [{\citenamefont {{Suzuki, T., Ikeda, S., Richardson, J.W.,
  Yamaguchi}}(1993)}]{Suzuki1993}%
  \BibitemOpen
  \bibfield  {author} {\bibinfo {author} {\bibfnamefont {Y.}~\bibnamefont
  {{Suzuki, T., Ikeda, S., Richardson, J.W., Yamaguchi}}},\ }in\ \href@noop {}
  {\emph {\bibinfo {booktitle} {Proceedings of the Fifth International
  Symposium on Advanced Nuclear Energy Research}}}\ (\bibinfo {year} {1993})\
  pp.\ \bibinfo {pages} {343--346}\BibitemShut {NoStop}%
\bibitem [{\citenamefont {Mankovsky}\ \emph {et~al.}(2016)\citenamefont
  {Mankovsky}, \citenamefont {Polesya}, \citenamefont {Ebert},\ and\
  \citenamefont {Bensch}}]{Mankovsky2016}%
  \BibitemOpen
  \bibfield  {author} {\bibinfo {author} {\bibfnamefont {S.}~\bibnamefont
  {Mankovsky}}, \bibinfo {author} {\bibfnamefont {S.}~\bibnamefont {Polesya}},
  \bibinfo {author} {\bibfnamefont {H.}~\bibnamefont {Ebert}}, \ and\ \bibinfo
  {author} {\bibfnamefont {W.}~\bibnamefont {Bensch}},\ }\href {\doibase
  10.1103/PhysRevB.94.184430} {\bibfield  {journal} {\bibinfo  {journal}
  {Physical Review B}\ }\textbf {\bibinfo {volume} {94}},\ \bibinfo {pages} {1}
  (\bibinfo {year} {2016})},\ \Eprint {http://arxiv.org/abs/1607.05738}
  {arXiv:1607.05738} \BibitemShut {NoStop}%
\bibitem [{\citenamefont {Wu}\ and\ \citenamefont {Birgeneau}(2021)}]{Wu2021}%
  \BibitemOpen
  \bibfield  {author} {\bibinfo {author} {\bibfnamefont {S.}~\bibnamefont
  {Wu}}\ and\ \bibinfo {author} {\bibfnamefont {R.~J.}\ \bibnamefont
  {Birgeneau}},\ }\href@noop {} {\bibfield  {journal} {\bibinfo  {journal}
  {unpublished}\ } (\bibinfo {year} {2021})}\BibitemShut {NoStop}%
\bibitem [{\citenamefont {Mahan}(2000)}]{Mahan2000}%
  \BibitemOpen
  \bibfield  {author} {\bibinfo {author} {\bibfnamefont {G.~D.}\ \bibnamefont
  {Mahan}},\ }\href@noop {} {\emph {\bibinfo {title} {{Many-Particle
  Physics}}}},\ \bibinfo {edition} {3rd}\ ed.\ (\bibinfo  {publisher} {Kluwer
  Academic/Plenum Publishers},\ \bibinfo {year} {2000})\BibitemShut {NoStop}%
\bibitem [{\citenamefont {Freimuth}\ \emph {et~al.}(2014)\citenamefont
  {Freimuth}, \citenamefont {Bl{\"{u}}gel},\ and\ \citenamefont
  {Mokrousov}}]{Freimuth2014}%
  \BibitemOpen
  \bibfield  {author} {\bibinfo {author} {\bibfnamefont {F.}~\bibnamefont
  {Freimuth}}, \bibinfo {author} {\bibfnamefont {S.}~\bibnamefont
  {Bl{\"{u}}gel}}, \ and\ \bibinfo {author} {\bibfnamefont {Y.}~\bibnamefont
  {Mokrousov}},\ }\href {\doibase 10.1103/PhysRevB.90.174423} {\bibfield
  {journal} {\bibinfo  {journal} {Physical Review B - Condensed Matter and
  Materials Physics}\ }\textbf {\bibinfo {volume} {90}},\ \bibinfo {pages} {1}
  (\bibinfo {year} {2014})},\ \Eprint {http://arxiv.org/abs/1305.4873}
  {arXiv:1305.4873} \BibitemShut {NoStop}%
\bibitem [{\citenamefont {{\v{Z}}elezn{\'{y}}}\ \emph
  {et~al.}(2017{\natexlab{b}})\citenamefont {{\v{Z}}elezn{\'{y}}},
  \citenamefont {Zhang}, \citenamefont {Felser},\ and\ \citenamefont
  {Yan}}]{Zelezny2017b}%
  \BibitemOpen
  \bibfield  {author} {\bibinfo {author} {\bibfnamefont {J.}~\bibnamefont
  {{\v{Z}}elezn{\'{y}}}}, \bibinfo {author} {\bibfnamefont {Y.}~\bibnamefont
  {Zhang}}, \bibinfo {author} {\bibfnamefont {C.}~\bibnamefont {Felser}}, \
  and\ \bibinfo {author} {\bibfnamefont {B.}~\bibnamefont {Yan}},\ }\href
  {\doibase 10.1103/PhysRevLett.119.187204} {\bibfield  {journal} {\bibinfo
  {journal} {Physical Review Letters}\ }\textbf {\bibinfo {volume} {119}},\
  \bibinfo {pages} {1} (\bibinfo {year} {2017}{\natexlab{b}})},\ \Eprint
  {http://arxiv.org/abs/1702.00295} {arXiv:1702.00295} \BibitemShut {NoStop}%
\bibitem [{\citenamefont {Zelezny}(2017)}]{Zelezny2017a}%
  \BibitemOpen
  \bibfield  {author} {\bibinfo {author} {\bibfnamefont {J.}~\bibnamefont
  {Zelezny}},\ }\href@noop {} {\enquote {\bibinfo {title} {{Linear Response
  Symmetry}},}\ } (\bibinfo {year} {2017})\BibitemShut {NoStop}%
\bibitem [{\citenamefont {Maniv}\ \emph {et~al.}(2021)\citenamefont {Maniv},
  \citenamefont {Nair}, \citenamefont {Haley}, \citenamefont {Doyle},
  \citenamefont {John}, \citenamefont {Cabrini}, \citenamefont {Maniv},
  \citenamefont {Ramakrishna}, \citenamefont {Tang}, \citenamefont {Ercius},
  \citenamefont {Ramesh}, \citenamefont {Tserkovnyak}, \citenamefont {Reyes},\
  and\ \citenamefont {Analytis}}]{Maniv2021}%
  \BibitemOpen
  \bibfield  {author} {\bibinfo {author} {\bibfnamefont {E.}~\bibnamefont
  {Maniv}}, \bibinfo {author} {\bibfnamefont {N.}~\bibnamefont {Nair}},
  \bibinfo {author} {\bibfnamefont {S.~C.}\ \bibnamefont {Haley}}, \bibinfo
  {author} {\bibfnamefont {S.}~\bibnamefont {Doyle}}, \bibinfo {author}
  {\bibfnamefont {C.}~\bibnamefont {John}}, \bibinfo {author} {\bibfnamefont
  {S.}~\bibnamefont {Cabrini}}, \bibinfo {author} {\bibfnamefont
  {A.}~\bibnamefont {Maniv}}, \bibinfo {author} {\bibfnamefont {S.~K.}\
  \bibnamefont {Ramakrishna}}, \bibinfo {author} {\bibfnamefont {Y.~L.}\
  \bibnamefont {Tang}}, \bibinfo {author} {\bibfnamefont {P.}~\bibnamefont
  {Ercius}}, \bibinfo {author} {\bibfnamefont {R.}~\bibnamefont {Ramesh}},
  \bibinfo {author} {\bibfnamefont {Y.}~\bibnamefont {Tserkovnyak}}, \bibinfo
  {author} {\bibfnamefont {A.~P.}\ \bibnamefont {Reyes}}, \ and\ \bibinfo
  {author} {\bibfnamefont {J.}~\bibnamefont {Analytis}},\ }\href@noop {}
  {\bibfield  {journal} {\bibinfo  {journal} {Science Advances}\ }\textbf
  {\bibinfo {volume} {7}},\ \bibinfo {pages} {1} (\bibinfo {year} {2021})},\
  \Eprint {http://arxiv.org/abs/2008.02795} {arXiv:2008.02795} \BibitemShut
  {NoStop}%
\bibitem [{\citenamefont {Maniv}(2020)}]{communManiv2020}%
  \BibitemOpen
  \bibfield  {author} {\bibinfo {author} {\bibfnamefont {E.}~\bibnamefont
  {Maniv}},\ }\href@noop {} {\bibfield  {journal} {\bibinfo  {journal} {private
  communication}\ } (\bibinfo {year} {2020})}\BibitemShut {NoStop}%
\bibitem [{\citenamefont {Zhang}\ \emph {et~al.}(2016)\citenamefont {Zhang},
  \citenamefont {Han}, \citenamefont {Yang}, \citenamefont {Sun}, \citenamefont
  {Zhang}, \citenamefont {Yan},\ and\ \citenamefont {Parkin}}]{Zhang2016}%
  \BibitemOpen
  \bibfield  {author} {\bibinfo {author} {\bibfnamefont {W.}~\bibnamefont
  {Zhang}}, \bibinfo {author} {\bibfnamefont {W.}~\bibnamefont {Han}}, \bibinfo
  {author} {\bibfnamefont {S.~H.}\ \bibnamefont {Yang}}, \bibinfo {author}
  {\bibfnamefont {Y.}~\bibnamefont {Sun}}, \bibinfo {author} {\bibfnamefont
  {Y.}~\bibnamefont {Zhang}}, \bibinfo {author} {\bibfnamefont
  {B.}~\bibnamefont {Yan}}, \ and\ \bibinfo {author} {\bibfnamefont {S.~S.}\
  \bibnamefont {Parkin}},\ }\href {\doibase 10.1126/sciadv.1600759} {\bibfield
  {journal} {\bibinfo  {journal} {Science Advances}\ }\textbf {\bibinfo
  {volume} {2}} (\bibinfo {year} {2016}),\ 10.1126/sciadv.1600759}\BibitemShut
  {NoStop}%
\bibitem [{\citenamefont {Kriegner}\ \emph {et~al.}(2016)\citenamefont
  {Kriegner}, \citenamefont {V{\'{y}}born{\'{y}}}, \citenamefont
  {Olejn{\'{i}}k}, \citenamefont {Reichlov{\'{a}}}, \citenamefont
  {Nov{\'{a}}k}, \citenamefont {Marti}, \citenamefont {Gazquez}, \citenamefont
  {Saidl}, \citenamefont {N{\v{e}}mec}, \citenamefont {Volobuev}, \citenamefont
  {Springholz}, \citenamefont {Hol{\'{y}}},\ and\ \citenamefont
  {Jungwirth}}]{Kriegner2016}%
  \BibitemOpen
  \bibfield  {author} {\bibinfo {author} {\bibfnamefont {D.}~\bibnamefont
  {Kriegner}}, \bibinfo {author} {\bibfnamefont {K.}~\bibnamefont
  {V{\'{y}}born{\'{y}}}}, \bibinfo {author} {\bibfnamefont {K.}~\bibnamefont
  {Olejn{\'{i}}k}}, \bibinfo {author} {\bibfnamefont {H.}~\bibnamefont
  {Reichlov{\'{a}}}}, \bibinfo {author} {\bibfnamefont {V.}~\bibnamefont
  {Nov{\'{a}}k}}, \bibinfo {author} {\bibfnamefont {X.}~\bibnamefont {Marti}},
  \bibinfo {author} {\bibfnamefont {J.}~\bibnamefont {Gazquez}}, \bibinfo
  {author} {\bibfnamefont {V.}~\bibnamefont {Saidl}}, \bibinfo {author}
  {\bibfnamefont {P.}~\bibnamefont {N{\v{e}}mec}}, \bibinfo {author}
  {\bibfnamefont {V.~V.}\ \bibnamefont {Volobuev}}, \bibinfo {author}
  {\bibfnamefont {G.}~\bibnamefont {Springholz}}, \bibinfo {author}
  {\bibfnamefont {V.}~\bibnamefont {Hol{\'{y}}}}, \ and\ \bibinfo {author}
  {\bibfnamefont {T.}~\bibnamefont {Jungwirth}},\ }\href {\doibase
  10.1038/ncomms11623} {\bibfield  {journal} {\bibinfo  {journal} {Nature
  Communications}\ }\textbf {\bibinfo {volume} {7}},\ \bibinfo {pages} {1}
  (\bibinfo {year} {2016})},\ \Eprint {http://arxiv.org/abs/1508.04877}
  {arXiv:1508.04877} \BibitemShut {NoStop}%
\bibitem [{\citenamefont {Pop{\v{c}}evi{\'{c}}}\ \emph
  {et~al.}(2020)\citenamefont {Pop{\v{c}}evi{\'{c}}}, \citenamefont
  {Batisti{\'{c}}}, \citenamefont {Smontara}, \citenamefont {Velebit},
  \citenamefont {Ja{\'{c}}imovi{\'{c}}}, \citenamefont {Martino}, \citenamefont
  {{\v{Z}}ivkovi{\'{c}}}, \citenamefont {Tsyrulin}, \citenamefont {Piatek},
  \citenamefont {Berger}, \citenamefont {Sidorenko}, \citenamefont {R{\o}nnow},
  \citenamefont {Bari{\v{s}}i{\'{c}}}, \citenamefont {Forr{\'{o}}},\ and\
  \citenamefont {Tuti{\v{s}}}}]{Popcevic2020}%
  \BibitemOpen
  \bibfield  {author} {\bibinfo {author} {\bibfnamefont {P.}~\bibnamefont
  {Pop{\v{c}}evi{\'{c}}}}, \bibinfo {author} {\bibfnamefont {I.}~\bibnamefont
  {Batisti{\'{c}}}}, \bibinfo {author} {\bibfnamefont {A.}~\bibnamefont
  {Smontara}}, \bibinfo {author} {\bibfnamefont {K.}~\bibnamefont {Velebit}},
  \bibinfo {author} {\bibfnamefont {J.}~\bibnamefont {Ja{\'{c}}imovi{\'{c}}}},
  \bibinfo {author} {\bibfnamefont {E.}~\bibnamefont {Martino}}, \bibinfo
  {author} {\bibfnamefont {I.}~\bibnamefont {{\v{Z}}ivkovi{\'{c}}}}, \bibinfo
  {author} {\bibfnamefont {N.}~\bibnamefont {Tsyrulin}}, \bibinfo {author}
  {\bibfnamefont {J.}~\bibnamefont {Piatek}}, \bibinfo {author} {\bibfnamefont
  {H.}~\bibnamefont {Berger}}, \bibinfo {author} {\bibfnamefont {A.~A.}\
  \bibnamefont {Sidorenko}}, \bibinfo {author} {\bibfnamefont {H.~M.}\
  \bibnamefont {R{\o}nnow}}, \bibinfo {author} {\bibfnamefont {N.}~\bibnamefont
  {Bari{\v{s}}i{\'{c}}}}, \bibinfo {author} {\bibfnamefont {L.}~\bibnamefont
  {Forr{\'{o}}}}, \ and\ \bibinfo {author} {\bibfnamefont {E.}~\bibnamefont
  {Tuti{\v{s}}}},\ }\href {http://arxiv.org/abs/2003.08127} {\  (\bibinfo
  {year} {2020})},\ \Eprint {http://arxiv.org/abs/2003.08127}
  {arXiv:2003.08127} \BibitemShut {NoStop}%
\end{thebibliography}%

\end{document}